\documentclass[aps,prd,preprint,preprintnumbers,showpacs]{revtex4-1}
\usepackage{graphicx}
\usepackage{amsmath}
\usepackage[section] {placeins}
\usepackage{multirow}
\usepackage{filecontents}
\begin{document}

\title{On $X(3872)$ production in high energy heavy ion  collisions}
\author{A.~Mart\'inez~Torres\footnote{amartine@if.usp.br}}

 \affiliation{
Instituto de F\'isica, Universidade de S\~ao Paulo, C.P. 66318, 05389-970 S\~ao 
Paulo, SP, Brazil.
}
\author{K.~P.~Khemchandani\footnote{kanchan@if.usp.br}}
 \affiliation{
Instituto de F\'isica, Universidade de S\~ao Paulo, C.P. 66318, 05389-970 S\~ao 
Paulo, SP, Brazil.
}
\author{ F.~S.~Navarra\footnote{navarra@if.usp.br}}
 \affiliation{
Instituto de F\'isica, Universidade de S\~ao Paulo, C.P. 66318, 05389-970 S\~ao 
Paulo, SP, Brazil.
}
\author{ M.~Nielsen\footnote{mnielsen@if.usp.br} }
 \affiliation{
Instituto de F\'isica, Universidade de S\~ao Paulo, C.P. 66318, 05389-970 S\~ao 
Paulo, SP, Brazil.
}
\author{ Luciano M. Abreu\footnote{luciano.abreu@ufba.br}}
\affiliation{Instituto de F\'isica, Universidade Federal da Bahia, 40210-340, Salvador, BA, Brazil.}
\preprint{}

\date{\today}

\begin{abstract}
We have determined the production cross sections of the $X(3872)$ state in the reactions $\bar D D\to\pi X$, $\bar D^* D\to \pi X$ and $\bar D^* D^*\to\pi X$, information which is useful for studies of the $X(3872)$ meson abundance in heavy ion collisions. We construct a formalism considering $X$ as a molecular bound state 
of $\bar D^0 D^{*0} - \textrm{c.c}$, $D^- D^{*+} - \textrm{c.c}$ and $D^-_s D^{*+}_s - \textrm{c.c}$. To obtain the amplitudes related to these processes we have made use of effective field Lagrangians. The evaluation  of the cross section of the processes involving $D^*$ meson(s) requires the calculation of
an anomalous vertex, $X\bar D^* D^*$, which has been obtained by considering triangular loops motivated by the molecular nature of $X(3872)$. Proceeding in this way, we have evaluated the cross section for the reaction $\bar D^* D\to \pi X$, and find that the diagrams involving the $X\bar D^* D^*$ vertex give a large contribution. Encouraged by this finding we estimate the $X\bar D^* D^*$ coupling, which turns out to be $1.95\pm 0.22$. We then use it to obtain the cross section for the reaction $\bar D^* D^*\to\pi X$ and find that, in this case too, the $X\bar D^* D^*$ vertex is relevant. We also discuss the role of the charged components of $X$ in the determination of the production cross sections.

\end{abstract}


\maketitle
 
\section{Introduction}

It is now a well accepted fact that in high energy heavy ion collisions a deconfined medium is created: 
the quark gluon plasma (QGP) \cite{Arsene,Adams}. Indeed, a significant part of 
the RHIC and LHC physics program is devoted to  determine  and understand the properties of the QGP. 
In another frontier of hadron physics, we find the $B$ factories BELLE \cite{belle} and BES \cite{bes}, which have produced a 
wealth of data on new hadronic states \cite{nora}. Particularly interesting are the data on the so called exotic charmonium 
states \cite{nora,review}. One member of this family, the $X(3872)$ (from now on simply $X$), was measured by many experimental 
groups and its existence is now established beyond any doubt. 

The first measurement of the $X$ meson was reported about a decade ago by the Belle collaboration~\cite{Choi}  
in the decay $B^{\pm}\to K^{\pm}\pi^+\pi^- J/\psi$ and it was subsequently confirmed by several other collaborations  
\cite{Acosta,Abazov,Aubert}, but only very recently the spin-parity quantum numbers of $X$ have been confirmed to be  
$1^{++}$~\cite{Aaij}. In this past decade since the discovery of the $X$, several theoretical models have been proposed  
for the structure of this new state, describing it as a charmonium state, a tetraquark, a $D - \bar{D^*}$ hadron molecule  
and a mixture between a charmonium and a molecular component~\cite{Tornqvist,Close,Swanson,Braaten,Daniel2,Matheus, 
Daniel3,Dong,Nielsen,Daniel4,Dubnicka,Badalian}. In spite of the effort of these numerous groups, the 
properties of this particle are not yet well understood and represent a challenge both for theorists and 
experimentalists.  

The ExHIC collaboration \cite{Cho1,Cho2} was created as a task force to investigate the fascinating possibility of 
learning more about exotic charmonium states in heavy ion collisions. In many aspects this program is a revival of 
the study of $J/\psi$ production in heavy ion collisions carried out fifteen years ago. The main difference is 
that in those days the $J/\psi$, a very well known charmonium state, was considered as a probe to understand 
the QGP. Now, in a remarkable inversion of strategy, we use the QGP to try to understand the new charmonium!
In a high energy heavy ion collision the QGP is formed, expands, cools, hadronizes and is converted into a 
hadron gas, which lives up to $10$ fm/c and then freezes out. Now as before, the most important part of this evolution is 
the QGP, where an increasing (with the reaction energy) number of charm quarks and anti-quarks move freely. The initially 
formed charmonium bound states are dissolved (the famous ``charmonium suppression") but $c$'s and $\bar{c}$'s, 
coming now from different parent 
gluons, can pick up  light quarks and anti-quarks from the rich environment and form multiquark bound states. This 
is called quark coalescence and it happens during the phase transition to the hadronic gas \cite{Cho1,Cho2}. 
Quark coalescence has proven to give a very successful description of particle production during the hadronization and can be 
applied to $X$ production from the plasma.

The formation of the quark gluon plasma phase increases the number of produced $X$'s \cite{Cho1,Cho2}. This fact alone 
is already stimulating for the study of the exotic charmonium. But there is more. The coalescence formalism is 
based on the overlap of the Wigner functions 
of the quarks and of the bound state, being thus sensitive to the spatial configuration of the charmonium state 
and hence being able to distinguish between a compact, $\simeq 1$ fm long, tetraquark configuration and a large 
$\simeq 10$ fm long, molecular configuration. A big difference between the predicted abundancies could be used as 
a tool to discriminate between different  $X$ structures and to help us to decide whether it is a molecule or a 
tetraquark \cite{Cho3}.  However (as before in the case of the $J/\psi$)  the long lasting hadron gas phase can change 
the yield coming from the plasma. The 
$X$'s can be destroyed in collisions with ordinary hadrons, such as $X + \pi \rightarrow D + \bar{D^*}$, and can also 
be produced through the inverse reactions, such as $D + \bar{D^*} \rightarrow X + \pi$.  We must then be able to 
calculate the cross sections of these processes. 

The theory of the interactions between charmonium and ordinary hadrons was developed to give a precise estimate of 
how strongly the charmonium is absorbed by a hadronic medium. Hadronic absorption was considered as a background 
for the most important suppression, which happened in the QGP, as a result of the color screening effect. This theory was 
based on effective Lagrangians with $SU(4)$ symmetry and it started to be developed in 1998, with the 
pioneering work of Matinyan and M\"uller \cite{mamu}. This work was  followed by successive improvements 
\cite{Oh,hag,linko,mane}, until 2003, when $J/\psi$ absorption cross sections were derived from QCD sum rules \cite{soma3}. 
In Ref.~\cite{Oh} it was shown that interaction terms with anomalous parity couplings  have a strong impact on the interaction 
cross section. Very recently in Ref.~\cite{Cho3}, the authors  revisited the subject, using this theory of charm meson 
interactions and including  vertices with the $X$. The interaction of the $X$ with other hadrons is essentially unknown.
The $X$  decays into $J/\psi \rho$ and into $J/\psi \omega$ and also into $D \bar{D^*}$. In the first theoretical works  
\cite{Maiani} addressing these decays, the required interaction Lagrangians were proposed for the  $X$-Vector-Vector ($X$VV) and 
$X$-Pseudoscalar-Vector ($X$PV) vertices. They were used in Ref.~\cite{Cho3}, where the hadronic absorption 
cross section of the $X$ by mesons like $\pi$ and $\rho$ was evaluated  
for the processes $\pi X \to D\bar D$, $\pi X \to D^*\bar D^*$,  $\rho X\to D\bar D$,  $\rho X \to D\bar D^*$, and 
$\rho X \to D^*\bar D^*$. Using these  cross sections, the variation of the $X$ meson abundance during the expansion of the 
hadronic matter was computed with the help of a kinetic equation with gain and loss terms. The results turned out to be  
strongly dependent on the quantum numbers of the $X$ and on its structure.

The present work is devoted to introduce two improvements in the calculation of cross sections performed in Ref.~\cite{Cho3}.  
The first and most important one is the inclusion of the anomalous vertices $\pi D^* D^*$ and $X \bar D^* D^*$, which were neglected 
before. With these vertices new reaction channels become possible, such as  $\pi X\to D\bar D^*$, and the inverse process
$D\bar D^* \to \pi X$. As will be seen, this reaction is the most important one for $X$ in the hadron gas. 
The relevance of anomalous couplings has also been shown earlier in different contexts, for example in the $J/\psi$ 
absorption cross sections by $\pi$ and $\rho$ mesons~\cite{Oh}, radiative decays of scalar resonances and axial vector 
mesons~\cite{Nagahiro1,Nagahiro2} and in kaon photoproduction~\cite{Ozaki}.

The second  improvement is the inclusions of the charged components of the $D$ and $D^*$ mesons which couple to the $X$.  
The fact that the mass of $X$ is very close to the $\bar D^0 D^{*0}$ threshold ($\sim 0.2$ MeV below it), while the 
charged components $D^- D^{*+}$ are bound by roughly 8 MeV, could make us think that the charged components might not play 
an important role in the description of the properties of $X$.  This is so because if a wave function is obtained for 
the neutral and charged components, the one associated with the neutral component, due to the small binding energy of the 
system, will extend much further away in space than the one related to the charged components. Thus, the former one has a 
larger probability to be found than the latter. This has often motivated an omission of the contribution of the charged 
components. See, for example,  Refs.~\cite{Swanson,Braaten}.  However, it was shown in Refs.~\cite{Daniel3,Daniel4} that 
the coupling of $X$ to the neutral and charged components is very similar. As argued in Refs.~\cite{Daniel3,Daniel4}, in 
strong processes, the relevant interactions are short ranged and it is the wave function at the origin that matters in 
the description of such processes. It was also shown in Ref.~\cite{Daniel4} that, for a molecular state formed due to 
the interaction of two hadrons, the wave function of the state at the origin is related to the coupling of this state 
with the hadrons constituting it. In the case of $X$, since the couplings to the neutral and charged open charm channels 
are found to be practically the same~\cite{Daniel3,Daniel4}, a good description of any short ranged process in which this 
state is involved would imply the contribution of both neutral and charged components. This fact is not incompatible with 
having a larger probability of finding the neutral components for $X$ when integrating the wave function over a large 
range~\cite{Daniel4}. 
In fact, the importance of the consideration of the neutral as well as the charged channels to describe the properties of 
$X$ has already been shown in calculations of decay widths of this state into $J/\psi\rho$, $J/\psi \omega$ and  
$J/\psi\gamma$, where differences of the order of a factor 20-30 were found for several branching ratios when the charged 
components were not included. Having in this case results not compatible with the experimental data on these branching 
ratios~\cite{Daniel4,Francesca2}.
 
We shall calculate the production cross sections of $X$ in the processes (a) $\bar D D \to \pi X$, 
(b) $\bar D^* D \to \pi X$ and (c) $\bar D^* D^*\to \pi X$. The determination of the latter two involves diagrams with the  
$\bar D^* D^*  X$ anomalous vertex. As will be seen, the consideration of this anomalous vertex is important and can not 
be neglected in the evaluation of the cross section of the processes (b) and (c). In order to calculate the cross sections, 
we consider the model of Refs.~\cite{Daniel3,Daniel4,Francesca2} in which $X$ is generated from the interaction of 
$\bar D^0 D^{*0}-\textrm{c.c}$, $D^- D^{*+}-\textrm{c.c}$ and $D^-_s D^{*+}_s-\textrm{c.c}$, thus, taking into account 
the neutral as well as the charged components.  To determine the cross section for the reaction (b) we 
consider triangular loops motivated by the molecular nature of $X$. Having done this, we determine the amplitude for 
the same reaction considering $X$ as an effective field and estimate the $X\bar D^* D^*$ coupling such that it reproduces 
the results obtained by calculating the triangular loops. Using this coupling, we determine the production cross section 
for the process $\bar D^* D^*\to \pi X$. Although this last quantity had already been calculated in Ref.~\cite{Cho3}, our result is 
more complete because it contains the anomalous couplings. 

The paper is organized as follows. In the next section we describe the formalism used to calculate the production 
cross section for the different reactions studied here considering triangular loops and with an effective Lagrangian 
to describe the $X\bar D^* D^*$ vertex. 
In Sec.~\ref{Res}, we show the importance of the charged components of $X$ as well as the anomalous vertex 
$X\bar D^* D^*$ and the results found for the cross sections $\bar D D, \bar D^* D\to \pi X$. 
We also show the results for the $\bar D^* D\to \pi X$ cross section with the estimated $X \bar D^* D^*$ coupling. 
Using the same coupling we calculate the production cross section for $\bar D^* D^*\to \pi X$. Finally, in 
Sec.~\ref{Sum} we draw some conclusions.

\section{Formalism}\label{For}
The isospin-spin averaged production cross section for the processes $\bar D D, \bar D^* D, \bar D^* D^*\to \pi X$, in the center of mas (CM) frame can be calculated as
\begin{align}
\sigma_r(s)=\frac{1}{16\pi\lambda(s,m^2_{1i,r},m^2_{2i,r})}\int^{t_{\textrm{max,r}}}_{t_\textrm{min,r}}dt\overline{\sum\limits_{\textrm{Isos},\textrm{spin}}}\left |\mathcal{M}_r(s,t)\right|^2,\label{cross}
\end{align}
where $r=1,2,3$ is an index associated with the reaction having $\bar D D$, $\bar D^* D$ and $\bar D^* D^*$ as initial states, respectively, $\sqrt{s}$ is the CM energy, and $m_{1i,r}$ and $m_{2i,r}$ represent the masses of the two
particles present in the initial state $i$ of the reaction $r$. As a convention, when considering the initial state of the reaction $r$, we relate the index 1 (2) to the particle with charm $-1$ ($+1$). In Eq.~(\ref{cross}), the function $\lambda(a,b,c)$ is the K\"alen function, $t_\textrm{min,r}$ and $t_{\textrm{max,r}}$ correspond to the minimum and maximum values, respectively, of the Mandelstam variable $t$ and $\mathcal{M}_r$ is the reduced matrix element for the process $r$.  The symbol $\overline{\sum\limits_{\textrm{spin},\textrm{Isos}}}$ represents the sum over the isospins and spins of the particles in the initial and final state, weighted by the isospin and spin degeneracy factors of the two particles forming the initial state for the reaction $r$, i.e.,
\begin{align}
\overline{\sum\limits_{\textrm{spin},\textrm{Isos}}}\left|\mathcal{M}_r\right|^2\to \frac{1}{(2I_{1i,r}+1)(2I_{2i,r}+1)}\frac{1}{(2s_{1i,r}+1)(2s_{2i,r}+1)}\sum\limits_{\textrm{spin},\textrm{Isos}}\left|\mathcal{M}_r\right|^2,
\end{align}
where,
\begin{align}
\sum\limits_{\textrm{spin},\textrm{Isos}}\left|\mathcal{M}_r\right|^2=\sum\limits_{Q_{1i},Q_{2i}}\left[\sum\limits_{\textrm{spin}}\left|\mathcal{M}^{(Q_{1i},Q_{2i})}_r\right|^2\right].\label{Mqq}
\end{align}
In Eq.~(\ref{Mqq}), $Q_{1i}$ and $Q_{2i}$ represent the charges for each of the two particles forming the initial state $i$ of the reaction $r$, which are combined to obtain total charge $Q_r=Q_{1i}+Q_{2i}=0,+1,-1$. In this way,
we have four possibilities: $(0,0)$, $(-,+)$, $(-,0)$ and $(0,+)$ and thus,
\begin{align}
\sum\limits_{\textrm{spin},\textrm{Isos}}\left|\mathcal{M}_r\right|^2=\sum\limits_{\textrm{spin}}\left(\left|\mathcal{M}^{(0,0)}_r\right|^2+\left|\mathcal{M}^{(-,+)}_r\right|^2+\left|\mathcal{M}^{(-,0)}_r\right|^2+\left|\mathcal{M}^{(0,+)}_r\right|^2\right).
\end{align}.

\begin{figure}[h!]
\includegraphics[width=0.75\textwidth]{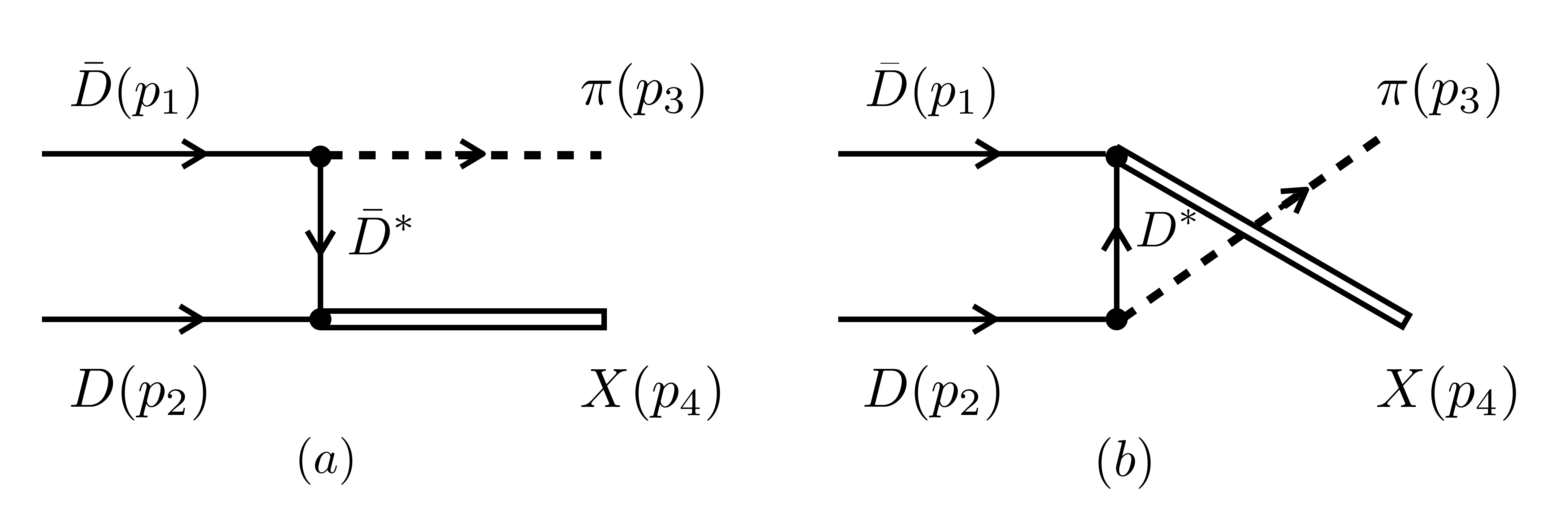}\\
\includegraphics[width=0.75\textwidth]{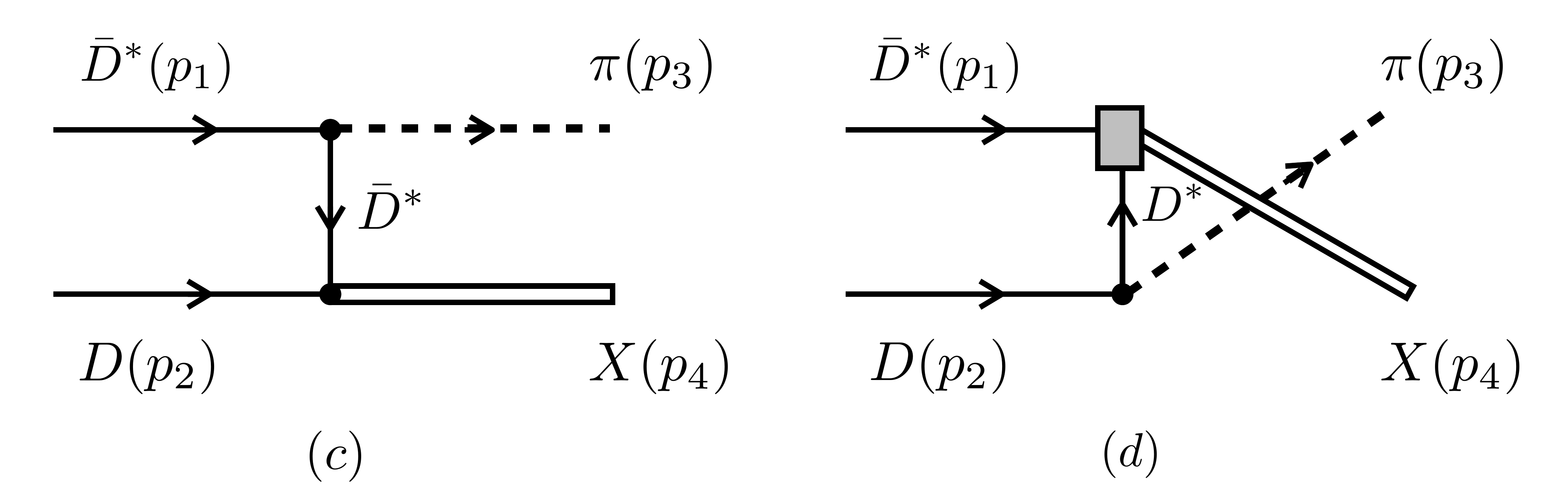}
\caption{Diagrams contributing to the process $\bar D D\to \pi X$ (top) and $\bar D^* D\to \pi X$ (bottom). The diagram containing a filled box is calculated by summing the set of diagrams shown in Fig.~\ref{blob}, as explained in the text.}\label{DbarstarD}
\end{figure}

\begin{figure}
\includegraphics[width=\textwidth]{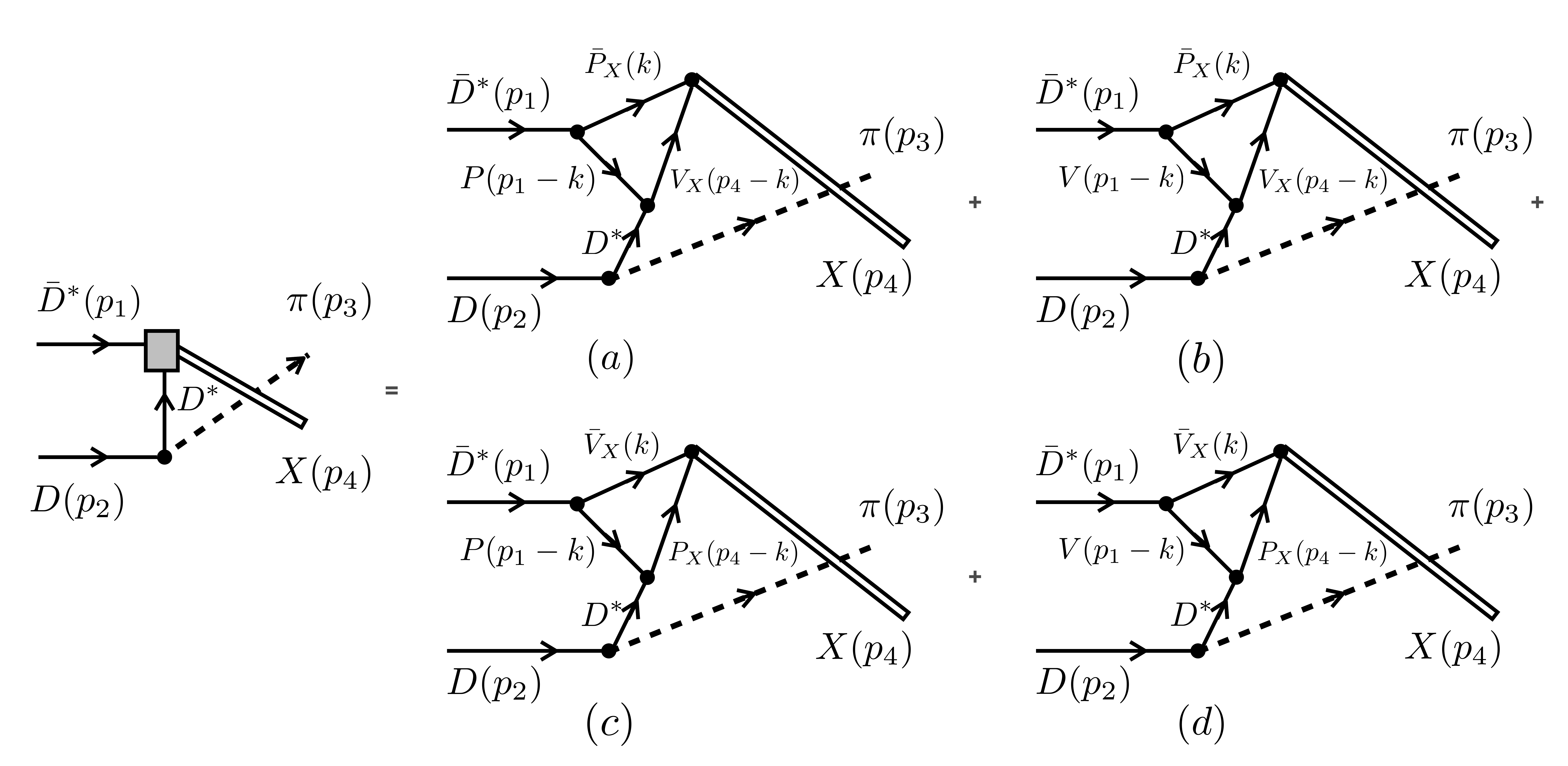}
\caption{Diagrams considered for the determination of the filled box shown in Fig.~\ref{DbarstarD}. The hadrons $P_X$ and $V_X$ represent the pseudoscalars and vectors coupling to the state $X$, while $P$ and $V$ are any
pseudoscalar and vector meson which can be exchanged conserving different quantum numbers. For a list of the different exchanged hadrons considered here see Table~\ref{ApenF3a}~and~\ref{ApenF3b} of the Appendix~\ref{AppenA}.}\label{blob}
\end{figure}

In Figs.~\ref{DbarstarD} and~\ref{blob} we show the different diagrams contributing to the processes $\bar D D\to \pi X$ and $\bar D^* D\to \pi X$ (without specifying the charge of the reaction).

Each of the amplitudes $\mathcal{M}^{(Q_{1i},Q_{2i})}_r$ of Eq.~(\ref{Mqq}) can be written as
\begin{align}
\mathcal{M}^{(Q_{1i},Q_{2i})}_r=T^{(Q_{1i},Q_{2i})}_r+U^{(Q_{1i},Q_{2i})}_r,
\end{align}
where $T^{(Q_{1i},Q_{2i})}_r$ and $U^{(Q_{1i},Q_{2i})}_r$ are the contributions related to the $t$ and $u$ channel diagrams shown in Fig.~\ref{DbarstarD} for the process $r$ with a charge $Q_r=Q_{1i}+Q_{2i}$.

To calculate the amplitudes for these $t$ and $u$ channel diagrams we need Lagrangians to determine the contribution of the Pseudoscalar-Pseudoscalar-Vector (PPV), Vector-Vector-Pseudoscalar  (VVP) and Vector-Vector-Vector (VVV) vertices. This can be done considering Lagrangians built using an effective theory in which the vector mesons are identified as the dynamical gauge bosons of the hidden $\textrm{U}(3)_V$ local symmetry in the $\textrm{U}(3)_L\times \textrm{U}(3)_R/\textrm{U}(3)_V$ non-linear sigma model~\cite{Bando1,Bando2,Meissner,Harada}, obtaining
\begin{align}
\mathcal{L}_{PPV}&=-i g_{PPV}  \langle V^\mu [P,\partial_\mu P]\rangle,\nonumber\\
\mathcal{L}_{VVP}&=\frac{g_{VVP}}{\sqrt{2}}\epsilon^{\mu\nu\alpha\beta}\langle\partial_\mu V_\nu\partial_\alpha V_\beta P\rangle\label{Lag}\\
\mathcal{L}_{VVV}&=i g_{VVV} \langle(V^\mu\partial_\nu V_\mu-\partial_\nu V_\mu V^\mu) V^\nu)\rangle.\nonumber
\end{align}
The $\mathcal{L}_{VVP}$ Lagrangian written above describes an anomalous vertex, which involves a violation of the natural parity. The natural parity of a particle 
is defined for bosons only and it is $P_n=P (-1)^J$, where $P$ is the intrinsic parity and $J$ is the spin of the 
particle. In other words, the natural parity of a particle is $+1$ if the particle transforms as a true Lorentz-tensor of that rank, and $-1$ if it transforms as a pseudotensor. In this way the field $V$ has natural parity $+1$, since it represents a vector, but the field $P$ has natural parity $-1$, since it corresponds to a pseudoscalar. There exists a unique way to construct the interaction Lagrangian that would violate the natural parity and would simultaneously conserve the intrinsic parity and would be Lorentz invariant: by using the Levi-Civita pseudotensor. So anomalous processes are described by a Lagrangian containing the Levi-Civita pseudotensor~\cite{WZ,Witten}.

The Lagrangians in Eq.~(\ref{Lag}) can be extended to SU(4) considering $P$ and $V_\mu$ as matrices containing the 15-plet of pseudoscalars and vectors mesons and the singlet of SU(4), respectively, which in the physical basis and considering ideal mixing for $\eta$ and $\eta^\prime$ as well as for $\omega$ and $\phi$  read as~\cite{Daniel}:
\begin{align}
P&=\left(\begin{array}{cccc}
\frac{\eta}{\sqrt{3}}+\frac{\eta^\prime}{\sqrt{6}}+\frac{\pi^0}{\sqrt{2}}&\pi^+&K^+&\bar{D}^0\\ 
\pi^- & \frac{\eta}{\sqrt{3}}+\frac{\eta^\prime}{\sqrt{6}}-\frac{\pi^0}{\sqrt{2}}&K^0&D^-\\
K^-&\bar{K}^0&-\frac{\eta}{\sqrt{3}}+\sqrt{\frac{2}{3}}\eta^\prime&D^-_s\\
D^0&D^+&D^+_s&\eta_c
\end{array}\right),
\end{align}
\begin{align}
V_\mu&=\left(\begin{array}{cccc}
\frac{\omega+\rho^0}{\sqrt{2}}&\rho^+&K^{*+}&\bar{D}^{*0}\\ 
\rho^- &\frac{\omega-\rho^0}{\sqrt{2}}&K^{*0}&D^{*-}\\
K^{*-}&\bar{K}^{*0}&\phi&D^{*-}_s\\
D^{*0}&D^{*+}&D^{*+}_s&J/\psi
\end{array}\right)_\mu.
\end{align}
The SU(4) symmetry is not a good symmetry in quantum cromodynamics, since the charm quark is much heavier than the $u$, $d$ and $s$ quarks. However it turns out that the SU(4) symmetry relations for couplings constants are not totally meaningless~\cite{Lin}. Nevertheless, the main idea of using the SU(4) symmetry here is to classify all the possible interaction vertices among the meson multiplets and then estimate their respective couplings trying to restrict them as much as possible by using available experimental information. 

In SU(3), the couplings appearing in Eq.~(\ref{Lag}) are given by~\cite{Nagahiro1,Francesca2,Khem}
\begin{align}
g_{PPV}&=\frac{m_V}{2 f_\pi},\quad
g_{VVP}=\frac{3 m^2_V}{16 \pi^2 f^3_\pi},\quad
g_{VVV}=\frac{m_V}{2 f_\pi},\label{coup}
\end{align}
with $m_V$ being the mass of the vector meson, which we take as the mass of the $\rho$ meson, and $f_\pi=93$ MeV is the pion decay constant.
The symbol $\langle\,\rangle$  in Eq.~(\ref{Lag}) indicates the trace in the isospin space. The coupling $g_{PPV}$  is the strong coupling of the $D^*$ meson to $D\pi$, however,
the value obtained from Eq.~(\ref{coup}) is $g_{PPV}\sim 4$, which is too small to reproduce the experimental decay width found for the process $D^*\to D\pi$. However, as
shown in Refs.~\cite{Liang,Francesca}, consideration of heavy quark symmetry gives an effective $g_{PPV}$ for the vertices involving $D$ and $D^*$ mesons as
\begin{align}
g_{PPV}=\frac{m_V}{2 f_\pi}\frac{m_{D^*}}{m_{K^*}}.\label{HQS}
\end{align}
With this additional factor, $g_{PPV}\sim 9$ and the decay width for the process $D^{*+}\to D^0\pi^+$ is 71 KeV, in agreement with the recent experimental result of $(65\pm 15)$ KeV~\cite{Anastassov}.
Thus we consider the coupling in Eq.~(\ref{HQS}) for the PPV Lagrangian, which is also compatible with the coupling found in Ref.~\cite{Bracco} using QCD sum rules.

As can be seen in Figs.~\ref{DbarstarD} and~\ref{blob}, the evaluation of the diagrams also requires the coupling of the $X$ state to the hadron components $\bar D D^* - \textrm{c.c}$, $\bar D_s D^{*}_s -\textrm{c.c}$. For this, we follow Refs.~\cite{Francesca2,Daniel2,Daniel3}, in which $X$ is generated from the dynamics of these hadrons, having a pole at $3871.6-i 0.001$ MeV with a coupling to the respective hadron components 
shown in Table~\ref{tableX}. 
\begin{table}
\caption{Couplings of $X$ to the different pseudoscalar-vector components constituting the state ($\bar{P}_X V_X$). The couplings for the complex conjugate components bear a minus sign.}
\label{tableX}
\begin{tabular}{c c}
\hline\hline
$\bar{P}_X V_X$&$g_{X\bar{P}_X V_X}$ (MeV)\\
\hline
$D^- D^{*+}$ & $3638/\sqrt{2}$ \\
$\bar{D}^0 D^{*0}$&$3663/\sqrt{2}$ \\
$D^-_s D^{*+}_s$ &$3395/\sqrt{2}$
\end{tabular}
\end{table}
As can be seen from Table~\ref{tableX}, the couplings of the state to the neutral and charged components are very similar (there is a very small isospin violation, less than $1\%$). The binding energy for the neutral $\bar D^0 D^{*0}-\textrm{c.c}$ component is around $0.2$ MeV, value much smaller
than the 8 MeV binding energy of the charged $D^- D^{*+}-\textrm{c.c}$ component. Intuitively, one might think that the $\bar D^0 D^{*0}-\textrm{c.c}$ component is the only relevant one, since the associated 
wave function extends much further than the one associated with the charged component. However, as shown in Ref.~\cite{Daniel4}, the relevant interactions in most processes are short ranged and then the wave functions around the origin, proportional to the couplings in the approach of Refs.~\cite{Francesca2,Daniel2,Daniel3}, are important. Thus, the wave function of $X$ is very close to the isospin $I=0$ combination of $\bar D^0 D^{*0}-\textrm{c.c}$ and $D^- D^{*+}-\textrm{c.c}$ and has a sizable fraction of  $D^-_s D^{*+}_s$. This approach has been very successful in describing different properties of $X$, as, for example, the decay widths of $X\to J/\psi \gamma, J/\psi \rho, J/\psi \omega$~\cite{Francesca2}, showing the importance of considering the neutral as well as the charged components of $X$ in the determination of these decay widths. This is the approach followed in the present paper.

The last element necessary for the calculation of the amplitudes shown in Fig.~\ref{DbarstarD} is the anomalous vertex $X\bar D^* D^*$.  A way to proceed, analogously to the one considered in Ref.~\cite{Maiani} to determine the coupling of $X$ to $J/\psi V$,
with $V$ a vector meson, is to construct an effective Lagrangian of the type
\begin{align}
\mathcal{L}_{X \bar D^* D^*}=i g_{X\bar D^* D^*} \epsilon^{\mu\nu\alpha\beta}\partial_\mu X_\nu \bar D^*_\alpha D^*_\beta,\label{effL}
\end{align}
and try to estimate somehow the unknown coupling $g_{X\bar D^* D^*}$. However, a model like this would lose its predictive power in the absence of any reasonable constrain on the value of the coupling  $g_{X\bar D^* D^*}$. The strategy followed in this paper consists of first determining the $\bar D^* D\to\pi X$ cross section by calculating the $X\bar D^* D^*$ vertex in terms of the loops shown in Fig.~\ref{blob}. After this is done, we obtain the cross section for the same process but using the Lagrangian in Eq.~(\ref{effL}) to evaluate the diagram in Fig.~\ref{DbarstarD}d and compare both results. In this way, we get a reliable estimation of the  $g_{X\bar D^* D^*}$ coupling.

Once all the ingredients needed for the evaluation of these amplitudes are defined, we can start writing the contribution of each diagram.  The $t$-channel amplitude for the process $\bar D D\to \pi X$ can be written as
\begin{align}
T^{(Q_{1i},Q_{2i})}_1=W^{(Q_{1i},Q_{2i})}_1\,g_{PPV}\,g_{X}\frac{1}{t-m^2_{\bar{D}^*}}\left[(p_1+p_3)_\mu+\frac{m^2_{\bar{D}}-m^2_\pi}{m^2_{\bar{D}^*}}p_{2\mu}\right]\epsilon^\mu_X(p_4),\label{Tampl1}
\end{align}
and for the process $\bar D^* D\to\pi X$ as
\begin{align}
T^{(Q_{1i},Q_{2i})}_2=W^{(Q_{1i},Q_{2i})}_2\,g_{VVP}\,g_{X}\frac{1}{t-m^2_{\bar{D}^*}}\epsilon^{\mu\nu\alpha\beta}p_{1\mu}p_{3\alpha}\epsilon_{\bar D^*\nu}(p_1)\epsilon_{X\beta}(p_4).\label{Tampl2}
\end{align}

In Eqs.~(\ref{Tampl1}) and (\ref{Tampl2}), $W^{(Q_{1i},Q_{2i})}_r$ ($r=1,2$) are isospin coefficients, listed in Table~\ref{Ccoeff}, $g_X$ is the coupling of $X$ to its hadron components, and it depends on the reaction and the charge $(Q_{1i},Q_{2i})$ configuration (note that we have not written explicitly this dependence of $g_X$ to simplify the notation), $m_{\bar {D}^*}$, $m_{\bar D}$, $m_\pi$ are average masses for the $\bar D^*$, $\bar D$ and $\pi$, $\epsilon_{\bar D^*} (p_1)$ and $\epsilon_X(p_4)$ are the polarization vectors of the $\bar D^*$ meson and $X$, respectively, and the greek letters indicate Lorentz indices.

\begin{table}[h!]
\caption{Coefficients $W^{(Q_{1i},Q_{2i})}_r$ and couplings $g_X$ for the amplitudes given in Eqs.~(\ref{Tampl1})~and~(\ref{Tampl2}). We have defined $g_n\equiv g_{X\bar D^0 D^{*0}}$ and $g_c\equiv g_{X D^- D^{*+}}$,
whose numerical values can be found in Table~\ref{tableX}.\\}\label{Ccoeff}
\begin{tabular}{cccc}
\hline\hline
$r$\quad\quad&$(Q_{1i},Q_{2i})$&$W_r$\quad\quad&$g_X$\\
\hline
\multirow{2}{*}{1}\quad\quad&$(0,0)$&$-1/\sqrt{2}$\quad\quad&$-g_n$\\&$(-,+)$&$1/\sqrt{2}$\quad\quad &$-g_c$\\
&$(-,0)$&$-1$\quad\quad&$-g_n$\\
&$(0,+)$&$-1$\quad\quad&$-g_c$\\
\hline
\multirow{2}{*}{2}\quad\quad&$(0,0)$&$-1/2$\quad\quad&$-g_n$\\&$(-,+)$&$1/2$\quad\quad &$-g_c$\\
&$(-,0)$&$-1/\sqrt{2}$\quad\quad&$-g_n$\\
&$(0,+)$&$-1/\sqrt{2}$\quad\quad&$-g_c$
\end{tabular}
\end{table}
For the reaction $\bar D D\to \pi X$, the $u$-channel amplitude of the diagram in Fig.~\ref{DbarstarD}b is written as
\begin{align}
U^{(Q_{1i},Q_{2i})}_1=Z^{(Q_{1i},Q_{2i})}g_{PPV} g_X\frac{1}{u-m^2_{D^*}}\left[(p_2+p_3)_\mu+\frac{m^2_D-m^2_\pi}{m^2_{D^*}}p_{1\mu}\right]\epsilon^\mu_X(p_4),\label{Uampl}
\end{align}
where the coefficients $Z^{(Q_{1i},Q_{2i})}$ and couplings $g_X$ are given in Table~\ref{Zcoeff}.

\begin{table}
\caption{Coefficients $Z^{(Q_{1i},Q_{2i})}$ and couplings $g_X$ for the amplitude given in Eq.~(\ref{Uampl}). We have defined $g_n\equiv g_{X\bar D^0 D^{*0}}$ and $g_c\equiv g_{XD^- D^{*+}}$,
whose numerical values can be found in Table~\ref{tableX}.\\}\label{Zcoeff}
\begin{tabular}{ccc}
\hline\hline
$(Q_{1i},Q_{2i})$&$Z_r$\quad\quad&$g_X$\\
\hline
$(0,0)$&$1/\sqrt{2}$\quad\quad&$g_n$\\
$(-,+)$&$-1/\sqrt{2}$\quad\quad &$g_c$\\
$(-,0)$&$1$\quad\quad&$g_c$\\
$(0,+)$&$1$\quad\quad&$g_n$\\
\end{tabular}
\end{table}
For the reaction $\bar D^* D\to \pi X$, the amplitude for the $u$-channel diagram shown in Fig.~\ref{DbarstarD}d can be calculated as
\begin{align}
U^{(Q_{1i},Q_{2i})}_2=\sum\limits_{p=a}^d U^{(Q_{1i},Q_{2i})}_{2p},
\end{align}
with $U^{(Q_{1i},Q_{2i})}_{2p}$ ($p=a,b,\cdots,d$) being the amplitudes associated with the diagrams depicted in Fig.~\ref{blob}. As can be seen, these amplitudes depend on the hadrons present in the triangular loops ($P$, $V_X$, etc.), since the couplings, propagators, etc., depend on them. Thus, to determine the amplitude of one of the diagram in Fig.~\ref{blob}, we need to evaluate the contribution from the possible intermediate states. For a list of the hadrons involved in these loops we refer the reader to the Appendix~\ref{AppenA}.
The final result for the amplitude of each diagram in Fig.~\ref{blob} can be obtained by summing over the amplitudes for the different intermediate states 
\begin{align}
U^{(Q_{1i},Q_{2i})}_{2p}=\sum\limits_{P,P_X, V_X,V}\mathcal{U}^{(Q_{1i},Q_{2i})}_{2p},
\end{align}
where $\mathcal{U}^{(Q_{1i},Q_{2i})}_{2p}$, $p=a,b,\textrm{etc.}$, is the amplitude for the diagram in Fig.~\ref{blob}p for a particular set of hadrons in the triangular loop. 
Using the Lagrangians describing the PPV, VVP and VVV vertices, we can determine
these amplitudes. Let us start with the diagram in Fig.~\ref{blob}a. Applying the Feynman rules we obtain:

\begin{align}
-i\, \mathcal{U}^{(Q_{1i},Q_{2i})}_{2a}=-\sqrt{2}\,g^2_{PPV} \,g_{VVP}\,g_{X\bar P_X V_X} F^{(Q_{1i},Q_{2i})}_A\frac{1}{u-m^2_{D*}}\epsilon^\mu_{\bar D^*}(p_1)\epsilon^{\mu^\prime\nu^\prime\alpha^\prime\beta^\prime}p_{2\mu^\prime}p_{3\nu^\prime}\epsilon_{X\beta^\prime}(p_4)\,\mathcal{I}_{\mu\alpha^\prime},\label{Ua}
\end{align}
 where
\begin{align}
\mathcal{I}_{\mu\alpha^\prime}\equiv\int\frac{d^4k}{(2\pi)^4}\frac{(2k-p_1)_\mu(p_4-k)_{\alpha^\prime}}{[k^2-m^2_{\bar P_X}+i\epsilon][(p_1-k)^2-m^2_P+i\epsilon][(p_4-k)^2-m^2_{V_X}+i\epsilon]}.\label{int}
\end{align}
In Eq.~(\ref{int}), $p_1$ and $p_4$ are the four momenta of the $\bar D^*$ and $X$ in the CM frame (see Fig.~\ref{blob}), $m_{\bar P_X}$ and $m_{V_X}$ the masses of the pseudoscalar-vector pair which couples to $X$ and $m_P$ the mass of the remaining pseudoscalar meson in the triangular loop (see Fig.~\ref{blob}). The coefficients $F^{(Q_{1i},Q_{2i})}_A$ for the different charge $(Q_{1i},Q_{2i})$ configurations and hadrons in the loop function of Fig.~\ref{blob}a can be found in Table~\ref{ApenF3a} of the Appendix~\ref{AppenA}.
To deduce to Eq.~(\ref{Ua}) we have made use of the antisymmetric properties of the Levi-Civita tensor and we have summed over the polarizations of the internal vector mesons using
\begin{align}
\sum \epsilon_{\bar D^*\mu}(p_2-p_3)\epsilon_{\bar D^*\nu}(p_2-p_3)&=-g_{\mu\nu}+\frac{(p_2-p_3)_\mu (p_2-p_3)_\nu}{m^2_{D^*}},\\
\sum \epsilon_{V_X\mu}(p_4-k)\epsilon_{V_X\nu}(p_4-k)&=-g_{\mu\nu}+\frac{(p_4-k)_\mu (p_4-k)_\nu}{m^2_X},
\end{align}
with $g_{\mu\nu}$ the metric tensor. For the evaluation of the integral $\mathcal{I}_{\mu\alpha^\prime}$ we refer the reader to the Appendix~\ref{AppenB}. As can be seen there, while the integration on the temporal part of the $k$ variable can be performed using Cauchy's theorem, the integration on the spatial part is logarithmically divergent and can be regularized using a cut-off of natural size, i.e.,$~\sim 1$ GeV, which corresponds to a reasonable average size of the hadrons~\cite{Nagahiro1,Raquel,Khem2}.

The evaluation of the amplitude related to the diagram in Fig.~\ref{blob}b is slightly different to the previous case, since it involves a VVV vertex. Using the Feynman rules, we find
\begin{align}
-i\, \mathcal{U}^{(Q_{1i},Q_{2i})}_{2b}&=-\frac{g_{PPV}\,g_{VVV}\,g_{VVP}\,g_{X\bar P_X V_X}}{\sqrt{2}}F^{(Q_{1i},Q_{2i})}_B\frac{1}{u-m^2_{D^*}}\epsilon_{\bar D^*\nu}(p_1)\epsilon^{\mu\nu\alpha\beta}p_{1\mu}\nonumber\\
&\quad\times\Big[\mathcal{P}_\beta\,\epsilon_{X\sigma}(p_4)\,\mathcal{H}^\sigma_\alpha+\epsilon_{X\beta}(p_4)\,\mathcal{P}^{\nu^\prime}\,\mathcal{J}_{\alpha\nu^\prime}],\label{Ub}
\end{align}
where
\begin{align}
\mathcal{P}_\mu&\equiv(p_2+p_3)_\mu-(p_2-p_3)_\mu\frac{m^2_D-m^2_\pi}{m^2_{D^*}},\nonumber\\
\mathcal{H}^\sigma_\alpha&\equiv\int\frac{d^4k}{(2\pi)^4}\frac{k_\alpha(2p_1-k)^\sigma}{[k^2-m^2_{\bar P_X}+i\epsilon][(p_1-k)^2-m^2_V+i\epsilon][(p_4-k)^2-m^2_{V_X}+i\epsilon]},\label{HJR}\\
\mathcal{J}_{\alpha\nu^\prime}&\equiv\int\frac{d^4k}{(2\pi)^4}\frac{k_\alpha(2k-p_1-p_4)_{\nu^\prime}}{[k^2-m^2_{\bar P_X}+i\epsilon][(p_1-k)^2-m^2_V+i\epsilon][(p_4-k)^2-m^2_{V_X}+i\epsilon]},\nonumber
\end{align}
and the coefficients $F^{(Q_{1i},Q_{2i})}_B$ can be found in Table~\ref{ApenF3b} of the Appendix~\ref{AppenA}. To obtain the expression given in Eq.~(\ref{Ub}), we have taken into account the fact that in the model of Refs.~\cite{Daniel2,Daniel3}, the
$X$ can be considered as a molecular state of $\bar D D^*-\textrm{c.c}$ and $\bar D_s D^*_s-\textrm{c.c}$, with its hadron components being in s-wave. In this case, the hadrons forming $X$ are nearly
on-shell and, thus, their respective momenta are negligible as compared to their energies. In such a situation, although the vector $V_X$ in diagram Fig.~\ref{blob}b is off-shell, the fact that the pseudoscalar
$\bar P_X$ interacts with the vector $V_X$ to generate $X$ implies that these two hadrons are not very far from being on-shell. In this case, we can approximate the sum over the polarizations of the vector $V_X$ by
\begin{align}
\sum \epsilon^\mu_{V_X}(p_4-k)\epsilon^\nu_{V_X}(p_4-k)\sim \delta^{ij},\label{deltaij}
\end{align}
with $i$ and $j$ spatial indices. However it would be more convenient to keep the covariant formalism instead of working with mixed indices (some spatial and other temporal-spatial).  For this, it is interesting to notice that the result of Eq.~(\ref{deltaij}) is always contracted with the polarization vector of $X$. Thus, the use of Eq.~(\ref{deltaij}) implies neglecting the temporal part of the polarization vector of $X$. This is appropriate when
determining the production cross sections near the threshold of the reaction (100-200 MeV above), as we do here. In this case, the $X$ meson is nearly at rest, thus its momentum is negligible as compared to its mass.
Therefore, if we use the approximation
\begin{align}
\sum \epsilon^\mu_{V_X}(p_4-k)\epsilon^\nu_{V_X}(p_4-k)\sim -g^{\mu\nu}.\label{gmunu}
\end{align}
instead of working with Eq.~(\ref{deltaij}), we would be including in the result a very small contribution arising from the temporal part of the polarization vector of $X$ (which is of order $|\vec{p}_4|/m_X$ and, for practical purposes, negligible) but we can keep the covariant formalism.
The expressions for $\mathcal{H}^\sigma_\alpha$, $\mathcal{J}_{\alpha\nu^\prime}$, $\mathcal{R}_\alpha$ can be found in the Appendix~\ref{AppenB}.

For the diagram in Fig.~\ref{blob}c, using the approximation in Eq.~(\ref{gmunu}) to sum over the polarizations of the internal vector meson $\bar V_X$, we have
\begin{align}
-i\, \mathcal{U}^{(Q_{1i},Q_{2i})}_{2c}&=-\frac{g^2_{PPV}\,g_{VVP}\,g_{XP_X \bar V_X}}{\sqrt{2}}F^{(Q_{1i},Q_{2i})}_C\frac{1}{u-m^2_{D^*}}\epsilon^{\mu\nu\alpha\beta}p_{1\mu}\epsilon_{\bar D^*\nu}(p_1)\epsilon_{X\beta}(p_4)\mathcal{P}^{\mu^\prime}\mathcal{R}_{\alpha\mu^\prime},\label{Uc}
\end{align}
where
\begin{align}
\mathcal{R}_{\alpha\mu^\prime}=\int\frac{d^4k}{(2\pi)^4}\frac{k_\alpha(p_1+p_4-2k)_{\mu^\prime}}{[k^2-m^2_{\bar V_X}+i\epsilon][(p_1-k)^2-m^2_P+i\epsilon][(p_4-k)^2-m^2_{P_X}+i\epsilon]}.\label{R}
\end{align}
The coefficients $F^{(Q_{1i},Q_{2i})}_C$ and the result for $\mathcal{R}_{\alpha\mu^\prime}$ can be found in the Appendices~\ref{AppenA} and ~\ref{AppenB}, respectively.

Similarly, for the diagram in Fig.~\ref{blob}d, considering the antisymmetric properties of the Levi-Civita tensor, the Lorentz condition and Eq.~(\ref{gmunu}) for the internal vector $\bar V_X$, we get
\begin{align}
-i\,\mathcal{U}^{(Q_{1i},Q_{2i})}_{2d}&=\frac{g_{PPV}\,g_{VVV}\,g_{VVP}\,g_{XP_X \bar V_X}}{\sqrt{2}}F^{(Q_{1i},Q_{2i})}_D\frac{1}{u-m^2_{D^*}}\epsilon^{\mu^\prime\nu^\prime\alpha^\prime\beta^\prime}\epsilon^\sigma_X(p_4)(p_2+p_3)_{\nu^\prime}(p_2-p_3)_{\mu^\prime}\nonumber\\
&\quad\times[2\mathcal{Q}_{\alpha^\prime\beta^\prime}\epsilon_{\bar D^*\sigma}(p_1)-2\mathcal{Q}_{\alpha^\prime\nu}\epsilon^\nu_{\bar D^*}(p_1)g_{\beta^\prime\sigma}-\mathcal{S}_{\alpha^\prime\sigma}\epsilon_{\bar D^*\beta^\prime}(p_1)],\label{Ud}
\end{align}
with
\begin{align}
\mathcal{Q}_{\alpha^\prime\beta^\prime}&=\int\frac{d^4k}{(2\pi)^4}\frac{(p_1-k)_{\alpha^\prime}k_{\beta^\prime}}{[k^2-m^2_{\bar V_X}+i\epsilon][(p_1-k)^2-m^2_V+i\epsilon][(p_4-k)^2-m^2_{P_X}+i\epsilon]},\label{Q}\\
\mathcal{S}_{\alpha^\prime\sigma}&=\int\frac{d^4k}{(2\pi)^4}\frac{(p_1-k)_{\alpha^\prime}(2p_1-k)_\sigma}{[k^2-m^2_{\bar V_X}+i\epsilon][(p_1-k)^2-m^2_V+i\epsilon][(p_4-k)^2-m^2_{P_X}+i\epsilon]}\label{S}.
\end{align}
The result for the coefficients $F^{(Q_{1i},Q_{2i})}_D$ and these integrals are given in the Appendices~\ref{AppenA} and ~\ref{AppenB}, respectively.

Now, if instead of considering the triangular loops of Figs.~\ref{blob} to determine the vertex $\bar D^* D\to X$, we use the Lagrangian in Eq.~(\ref{effL}), we obtain for
the $u$-channel diagram in Fig.~\ref{DbarstarD}d the following amplitude
\begin{align}
U^{(Q_{1i},Q_{2i})}_2&=Z^{(Q_{1i},Q_{2i})}g_{PPV} g_{X\bar D^* D^*}\frac{1}{u-m^2_{D^*}}\epsilon^{\mu\nu\alpha\beta}p_{4\mu}\left[(p_2+p_3)_\alpha+\frac{m^2_D-m^2_\pi}{m^2_{D^*}}p_{1\alpha}\right]\nonumber\\
&\quad\times\epsilon_{X\nu}(p_4)\epsilon_{\bar D^*\beta}(p_1),\label{Uampl2}
\end{align}
where $g_{X\bar D^* D^*}$ is the coupling of $X$ to $\bar D^* D^*$. Since the isospin 0 combination of $\bar D^* D^*$ is proportional to $|\bar D^{*0} D^{*0}+D^{*-}D^{*+}\rangle$, we have that $g_{X\bar D^{*0} D^{*0}}=g_{X D^{*-} D^{*+}}\equiv g_{X\bar D^* D^*}$. The coefficients $Z^{(Q_{1i},Q_{2i})}$ are given in Table~\ref{Zcoeff}.

Using the amplitudes written above, we can determine the contribution from the $t$ and $u$ channel diagrams in Figs.~\ref{DbarstarD} and \ref{blob} and obtained the amplitude $\mathcal{M}_r$ needed to calculate the cross sections.

\section{Results}\label{Res}
\subsection{The $\bar D D\to \pi X$ reaction}
\begin{figure}[t]
\includegraphics[width=0.5\textwidth]{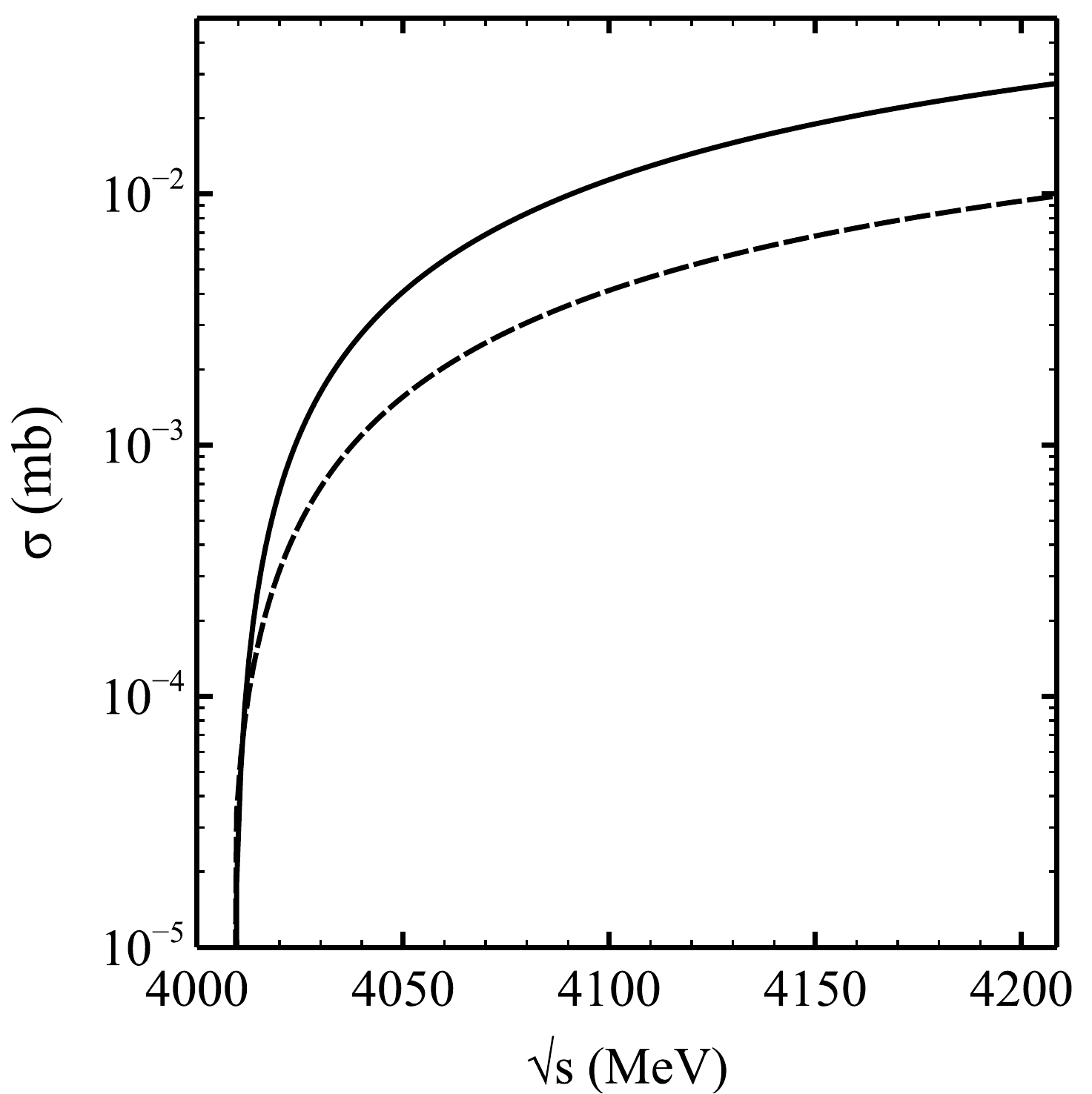}
\caption{Cross section for the reaction $\bar D D\to\pi X$ considering only the neutral components of X (dashed-line) and adding the charged components (solid line).}\label{crossDbarD}
\end{figure}

 In Fig.~\ref{crossDbarD} we show the results obtained for the production cross section of $X$ from the reaction $\bar D D\to\pi X$ as a function of the center of mass energy, $\sqrt{s}$. The dashed line corresponds to the case where only the neutral
components of $X$, i.e, $\bar D^0 D^{*0}-\textrm{c.c}$, are considered in the calculations, as in Ref.~\cite{Cho3}. The solid line is the result for the cross section when all components of $X$ are taken into account (using the couplings shown in Table~\ref{tableX}).
As can be seen from Fig.~\ref{crossDbarD}, the difference between the two curves is around a factor 2-3, depending on the energy. Thus, in a model in which $X$ is considered as a molecular state of $\bar D D^*-\textrm{c.c}$, a precise determination of the magnitude of the production cross section for $X$ necessarily implies the consideration of all the components, neutral as well as charged.

\begin{figure}
\includegraphics[width=0.5\textwidth]{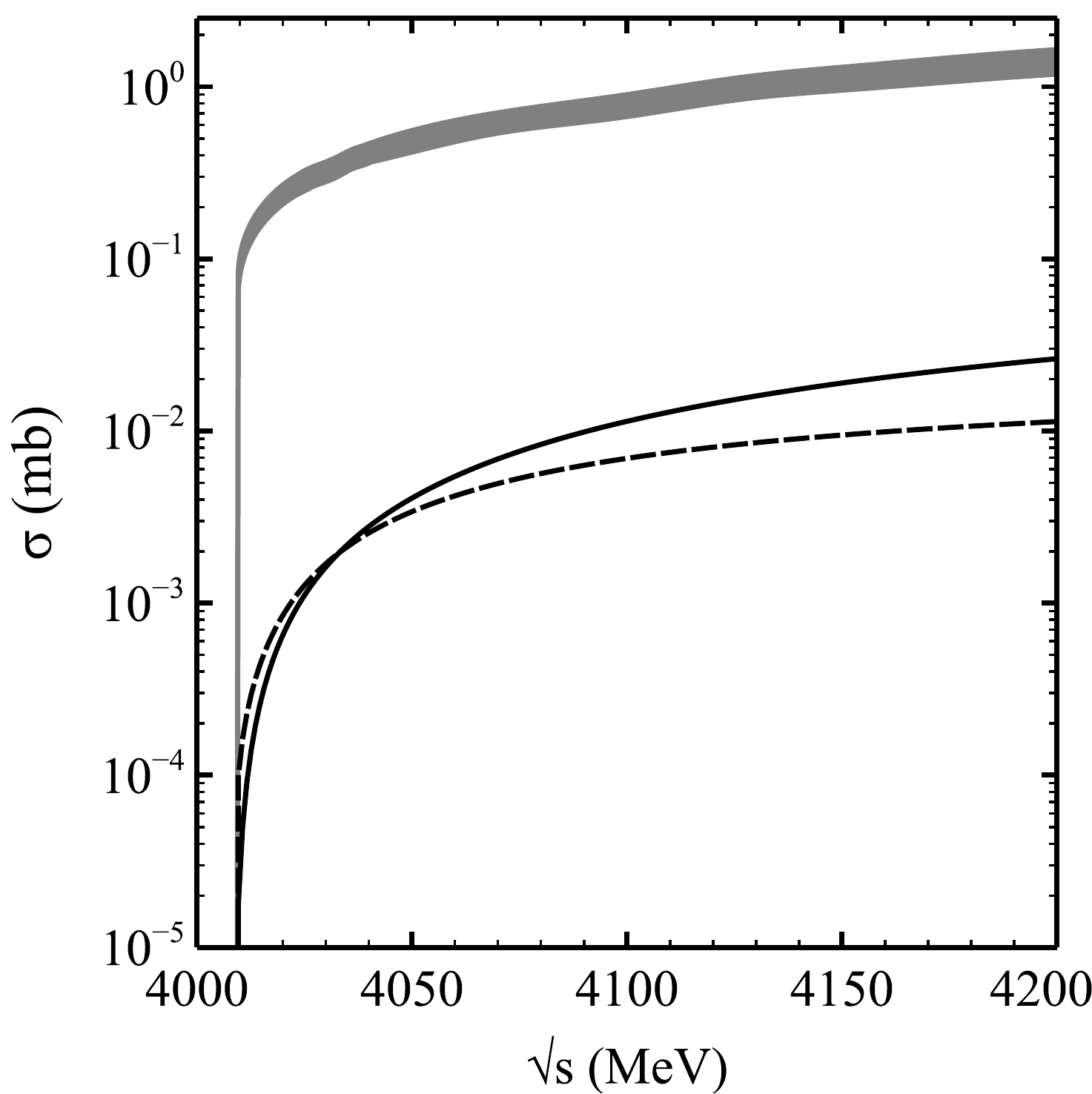}
\caption{Cross section for the reaction $\bar D^* D\to\pi X$. The solid line has the same meaning as in Fig.~\ref{crossDbarD}, and we have shown it for the purpose of comparison. The dashed line represents the result for the cross section of the process $\bar D^* D\to\pi X$ considering only the $t$ channel diagram in Fig.~\ref{DbarstarD}. The shaded region is the result obtained with both $t$ and $u$ channel diagrams of Fig.~\ref{DbarstarD} considering cut-offs in the range 700-1000 MeV.}\label{crossDbarstarD}
\end{figure}

\subsection{The $\bar D^* D\to \pi X$ reaction considering triangular loops}
Next, we determine the cross section related to the process $\bar D^* D\to \pi X$. The diagrams considered for this process (see Figs.~\ref{DbarstarD}c and~\ref{DbarstarD}d) involve anomalous vertices, $\bar D^* \bar D^*\pi$ in the $t$-channel and $X\bar D^* D^*$ in the $u$-channel. We find it interesting to compare the contributions arising form these vertices. We show the results in
Fig.~\ref{crossDbarstarD}. The solid line, as in Fig.~\ref{crossDbarD}, continues representing the final result for the $\bar D D\to \pi X$ cross section. The dashed line is the cross section for the $\bar D^* D\to \pi X$ process without considering the diagrams involving the anomalous vertex $X\bar D^* D^*$, i.e., only with the $t$ channel diagram shown in Fig.~\ref{DbarstarD}c. The shaded region represents the result found with both $t$ and $u$ channel diagrams shown in Figs.~\ref{DbarstarD}c and ~\ref{blob} (with the latter ones involving the $X\bar D^* D^*$ vertex) when changing the cut-off needed to regularize the loop integrals in the range 700-1000 MeV. As can be seen, the results do not get very affected by a reasonable change in the cut-off. Clearly, the vertex $X\bar D^* D^*$ plays an important role in the determination of the $\bar D^* D\to \pi X$ cross section, raising it by around a factor 100-150.

The importance of the anomalous vertices has been earlier mentioned in different contexts. For example, in Ref.~\cite{Oh} the $J/\psi$ absorption cross sections by $\pi$ and $\rho$ mesons were evaluated for several processes producing $D$ and $D^*$ mesons in the final state. The authors found
that  the $J/\psi\,\pi\to  D^*\bar D$ cross section obtained with the exchange of a $D^*$ meson in the $t$-channel, which involves the anomalous $D^* D^*\pi$ coupling, was around 80 times bigger than the one obtained with a $D$ meson exchange in the $t$-channel. In Ref.~\cite{Nagahiro1} the authors studied the radiative decay modes of the $f_0(980)$ and $a_0(980)$ resonances, finding
that the diagrams involving anomalous couplings were quite important for most of the decays, particularly for the $f_0(980)\to \rho^0\gamma$, $a_0(980)\to \rho\gamma$ and $a_0(980)\to\omega\gamma$.

Summarizing this subsection, we have shown that the cross section for the reaction $\bar D^* D\to\pi X$ is larger than that for $\bar D D\to\pi X$ and, thus, the consideration of this reaction in a calculation of the abundance of the $X$ meson in heavy ion collisions could be important.

\subsection{Estimating the $g_{X\bar D^* D^*}$ coupling}

\begin{figure}
\includegraphics[width=0.5\textwidth]{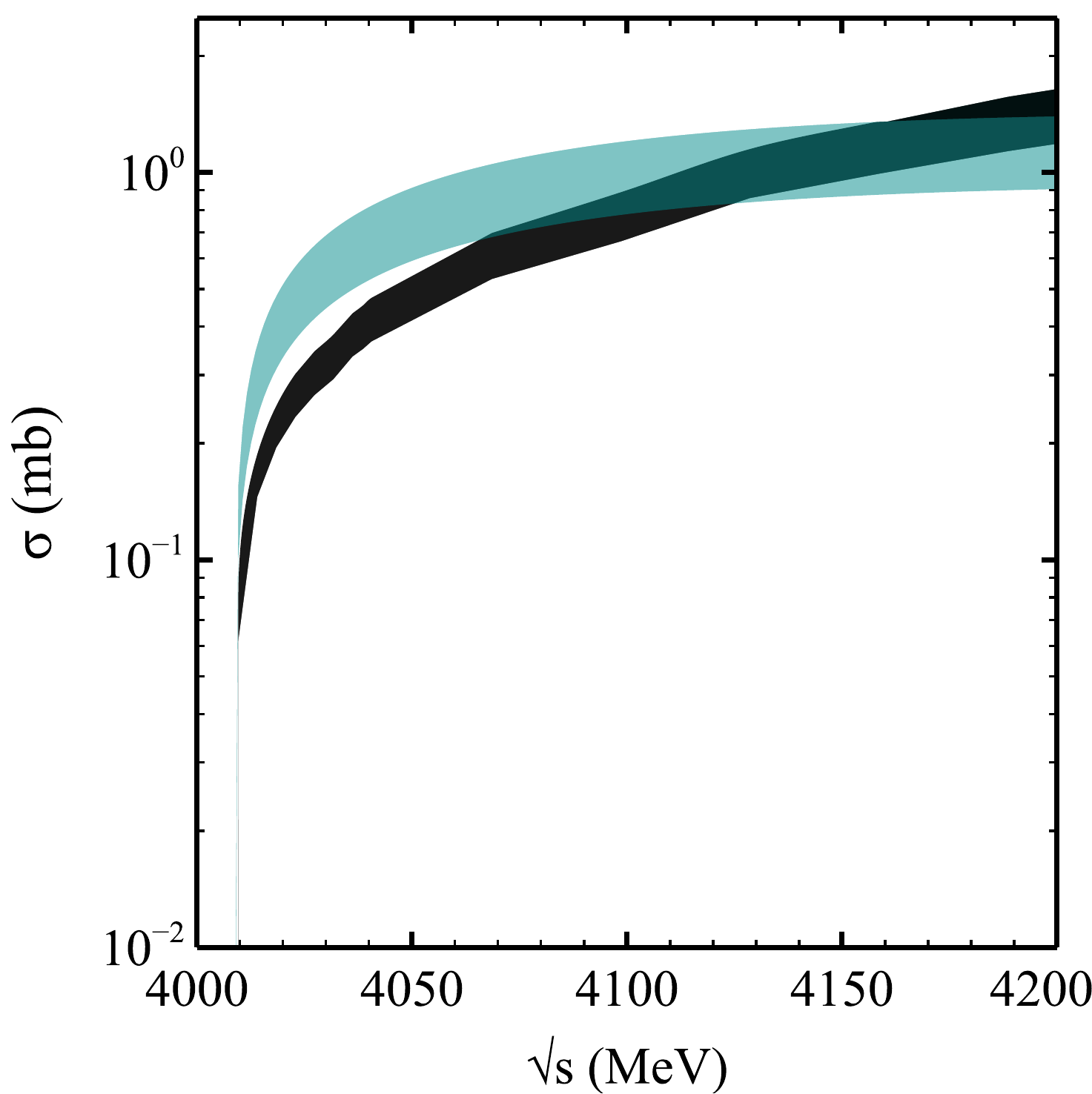}
\caption{Cross section for the reaction $\bar D^* D\to\pi X$. The dark color shaded region has the same meaning as the shaded region in Fig.~\ref{crossDbarstarD}. The light color shaded region represents the result for the cross section when considering the Lagrangian in Eq.~(\ref{effL}) to determine the $X\bar D^* D^*$ vertex with the value of the coupling given in Eq.~(\ref{coupa}).}\label{comp}
\end{figure}

Having determined the contribution from the anomalous vertex $X\bar D^* D^*$ calculating the loops shown in Fig.~\ref{blob}, we could now obtain the cross section for the $\bar D^* D\to\pi X$ reaction using the Lagrangian of Eq.~(\ref{effL}) to determine the amplitude for the diagram shown in Fig.~\ref{DbarstarD}d, which results in Eq.~(\ref{Uampl2}). In this way we can fix the $X\bar D^* D^*$ coupling to that value which gives similar results to the shaded region shown in Fig.~\ref{crossDbarstarD}. From Eq.~(\ref{effL}), it can be seen that the coupling $g_{X\bar D^* D^*}$ should be dimensionless. In Fig.~\ref{comp} we show the results found for the cross section of the reaction $\bar D^* D\to\pi X$ for  $g_{X\bar D^* D^*}$ in the range $1.95\pm 0.22$ (light color shaded region). The dark shaded region in the figure corresponds to the result for the cross section obtained by evaluating the vertex $X\bar D^* D^*$ using the diagrams in Fig.~\ref{blob}, where the loops have been regularized with a cut-off in the range $700-1000$ MeV. It can be seen that, although the energy dependence obtained by using the Lagrangian in Eq.~(\ref{effL}) is not exactly the same as the one found by  considering the triangular loops of Fig.~\ref{blob}, the two results are compatible in some energy range. Thus, the usage of the Lagrangian of Eq.~(\ref{effL}) with the value 
\begin{equation}
g_{X\bar D^* D^*}\sim 1.95\pm 0.22,\label{coupa}
\end{equation}
can be considered as a reasonable approximation for describing process involving the anomalous vertex $X\bar D^* D^*$, simplifying in this way the calculation of this vertex to a great extent.
\subsection{The $\bar D^* D^*\to\pi X$ reaction}
\begin{figure}
\includegraphics[width=0.75\textwidth]{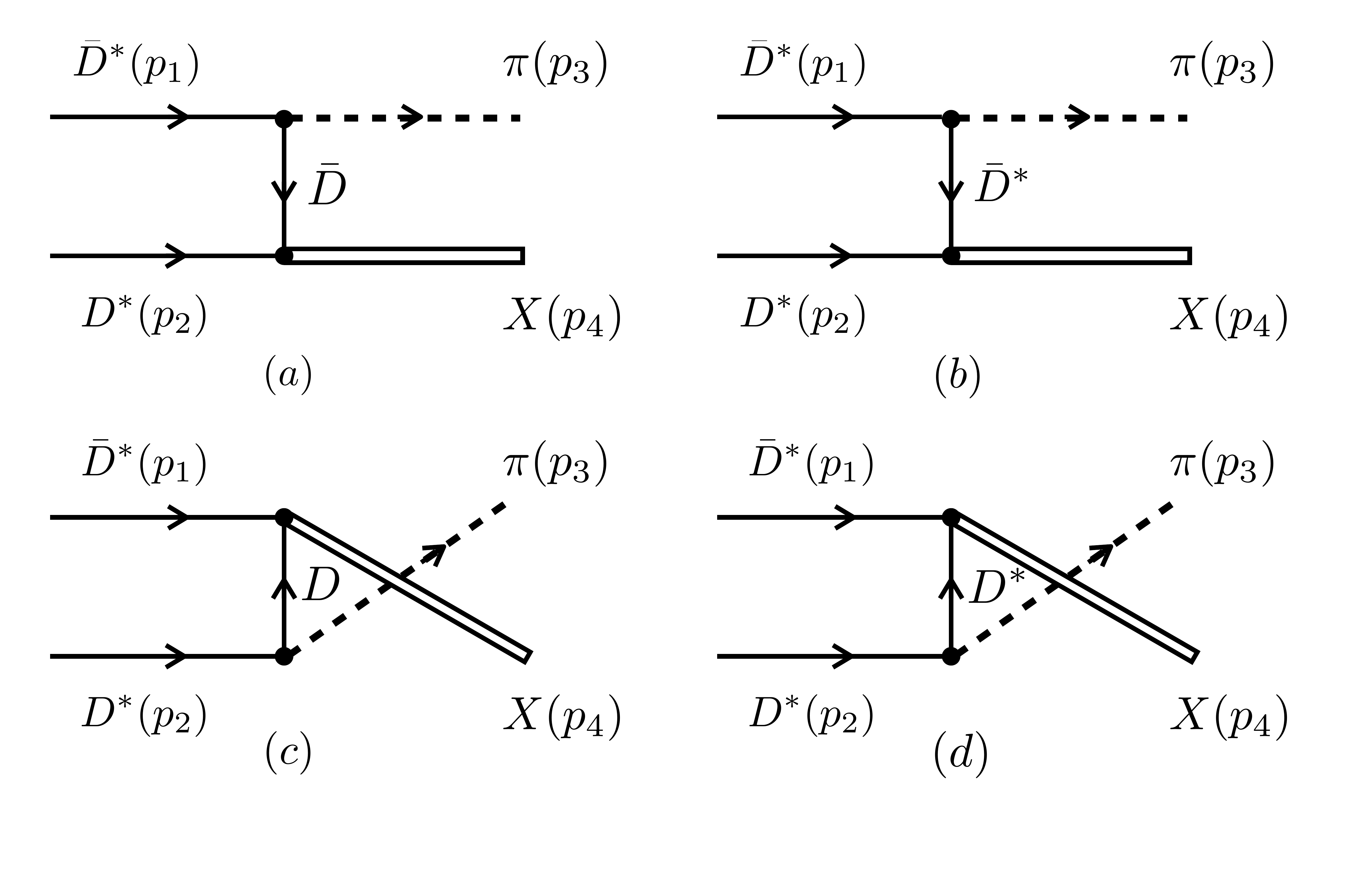}
\caption{Different diagrams contributing to the reaction $\bar D^* D^*\to\pi X$.}\label{DstarDstar}
\end{figure}

After estimating the coupling $g_{X\bar D^* D^*}$, we can use this value to determine the cross section for the process 
 $\bar D^* D^*\to\pi X$, which could also get a contribution from the anomalous $X\bar D^* D^*$ vertex, 
that was neglected in Ref.~\cite{Cho3}. The different Feynman diagrams considered for this process are depicted in Fig.~\ref{DstarDstar}.

\begin{figure}
\includegraphics[width=0.5\textwidth]{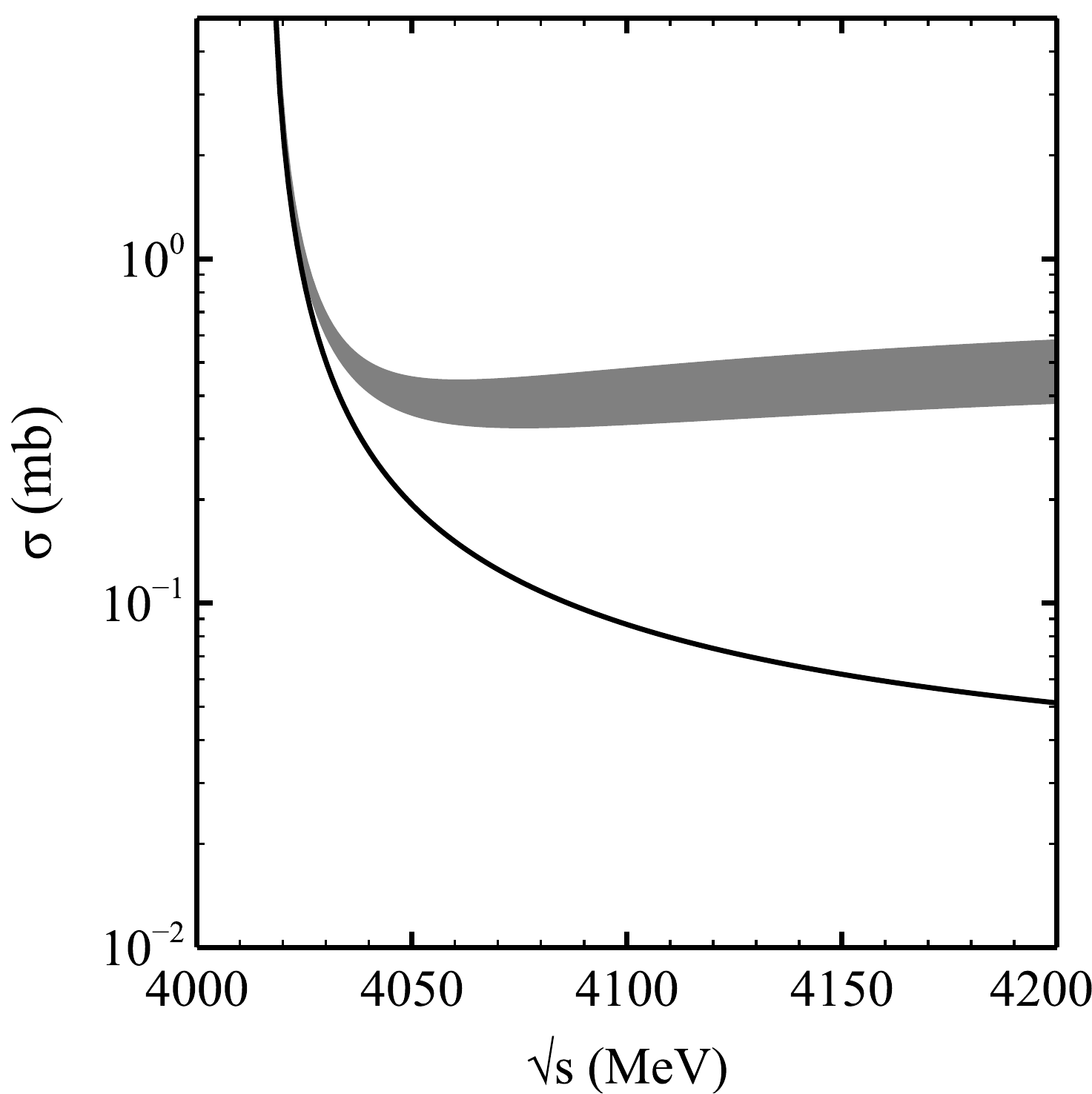}
\caption{Cross section for the reaction $\bar D^* D^*\to\pi X$. The solid line represents the cross section without the contribution from the diagrams in Fig.~\ref{DstarDstar}b and \ref{DstarDstar}d, which contain the vertex $X\bar D^* D^*$. The shaded region represents the result for the cross section when including the contribution of all the diagrams in Fig.~\ref{DstarDstar}, with the vertex $X\bar D^*D^*$ obtained using the Lagrangian in Eq.~(\ref{effL}) with the value of the coupling given in Eq.~(\ref{coupa}).}\label{figDbarstar}
\end{figure}

Considering the Lagrangian in Eq.~(\ref{effL}) for the $X\bar D^* D^*$ vertex, we find the following amplitudes for the $t$ and $u$ channel diagrams:
\begin{align}
T^{(Q_{1i},Q_{2i})}_{3a}&=-2 g_{PPV}\,g^a_{X}\,\mathcal{Y}^{(Q_{1i},Q_{2i})}\frac{1}{t-m^2_{\bar D}}p_{3\mu}\epsilon^\mu_{\bar D^*}(p_1)\epsilon^\nu_{D^*}(p_2)\epsilon_{X\nu}(p_4)\nonumber\\
T^{(Q_{1i},Q_{2i})}_{3b}&=-\frac{g_{VVP}}{\sqrt{2}}g_{X\bar D^* D^*}\,\mathcal{Y}^{(Q_{1i},Q_{2i})}\frac{1}{t-m^2_{\bar D^*}}\,\epsilon^{\mu\nu\alpha\beta}\,\epsilon^{\mu^\prime\nu^\prime\alpha^\prime}_{\phantom{\alpha}\phantom{\mu}\phantom{\nu}\phantom{\beta}\beta} \,p_{1\mu}\,p_{3\alpha}\,p_{4\mu^\prime}\epsilon_{\bar D^*\nu}(p_1)\epsilon_{D^*\alpha^\prime}(p_2)\epsilon_{X\nu^\prime}(p_4)\label{TUDstar}\\
U^{(Q_{1i},Q_{2i})}_{3c}&=-2 g_{PPV}\,g^c_{X}\,\mathcal{Y}^{(Q_{1i},Q_{2i})}\frac{1}{u-m^2_{D}}\,p_{3\nu}\,\epsilon^\mu_{\bar D^*}(p_1)\epsilon^\nu_{D^*}(p_2)\epsilon_{X\mu}(p_4)\nonumber\\
U^{(Q_{1i},Q_{2i})}_{3d}&=- \frac{g_{VVP}}{\sqrt{2}}g_{X\bar D^* D^*}\,\mathcal{Y}^{(Q_{1i},Q_{2i})}\frac{1}{u-m^2_{D^*}}\epsilon^{\mu\nu\alpha\beta}\epsilon^{\mu^\prime\nu^\prime\alpha^\prime\beta^\prime} g_{\nu^\prime\alpha}p_{2\alpha^\prime}p_{3\mu^\prime}p_{4\mu}\epsilon_{\bar D^*\beta}(p_1)\epsilon_{D^*\beta^\prime}(p_2)\epsilon_{X\nu}(p_4),\nonumber
\end{align}
where the values of $g^a_X$, $g^c_X$, and $\mathcal{Y}^{(Q_{1i},Q_{2i})}$ are those given in Table~\ref{Ycoeff}.
\begin{table}
\caption{Values for the coupling $g^{a,c}_X$ and the coefficients $\mathcal{Y}^{(Q_{1i},Q_{2i})}$ of Eq.~(\ref{TUDstar}). The numerical values of $g_n$ and $g_c$ can be found in Table~\ref{tableX}.}\label{Ycoeff}
\begin{tabular}{cccc}
\hline\hline
$(Q_{1i},Q_{2i})$&$g^a_X$&$g^c_X$&$\mathcal{Y}^{(Q_{1i},Q_{2i})}$\\
\hline
$(0,0)$&$g_n$&$g_n$&$\frac{1}{\sqrt{2}}$\\
$(-,+)$&$g_c$&$g_c$&$-\frac{1}{\sqrt{2}}$\\
$(-,0)$&$g_n$&$g_c$&$1$\\
$(0,+)$&$g_c$&$g_n$&$1$
\end{tabular}
\end{table}
In Fig.~\ref{figDbarstar} we show the results for the cross section of the reaction $\bar D^* D^*\to \pi X$. The solid line corresponds to the result found without the anomalous $X\bar D^* D^*$ contribution, while the shaded region is the result considering the diagrams involving this anomalous vertex with the value for the $g_{X\bar D^* D^*}$ coupling given in Eq.~(\ref{coupa}).
The first observation to be made is that the cross section for $\bar D^* D^*\to \pi X$ diverges close to the threshold of the reaction. This behavior is different to the cross sections of the processes studied in the previous sections. This is because the reaction $\bar D^* D^*\to \pi X$ is exothermic, while $\bar D D, \bar D^* D\to \pi X$ are endothermic. The second observation is that the contribution from the diagrams involving the $X\bar D^* D^*$ vertex is important, raising the cross section about a factor 8-10. 

Therefore, as in case of the $\bar D^* D\to\pi X$ reaction, the consideration of the anomalous vertices could play an important role when determining the $X$ abundance in heavy ion collisions.

\begin{figure}
\includegraphics[width=0.5\textwidth]{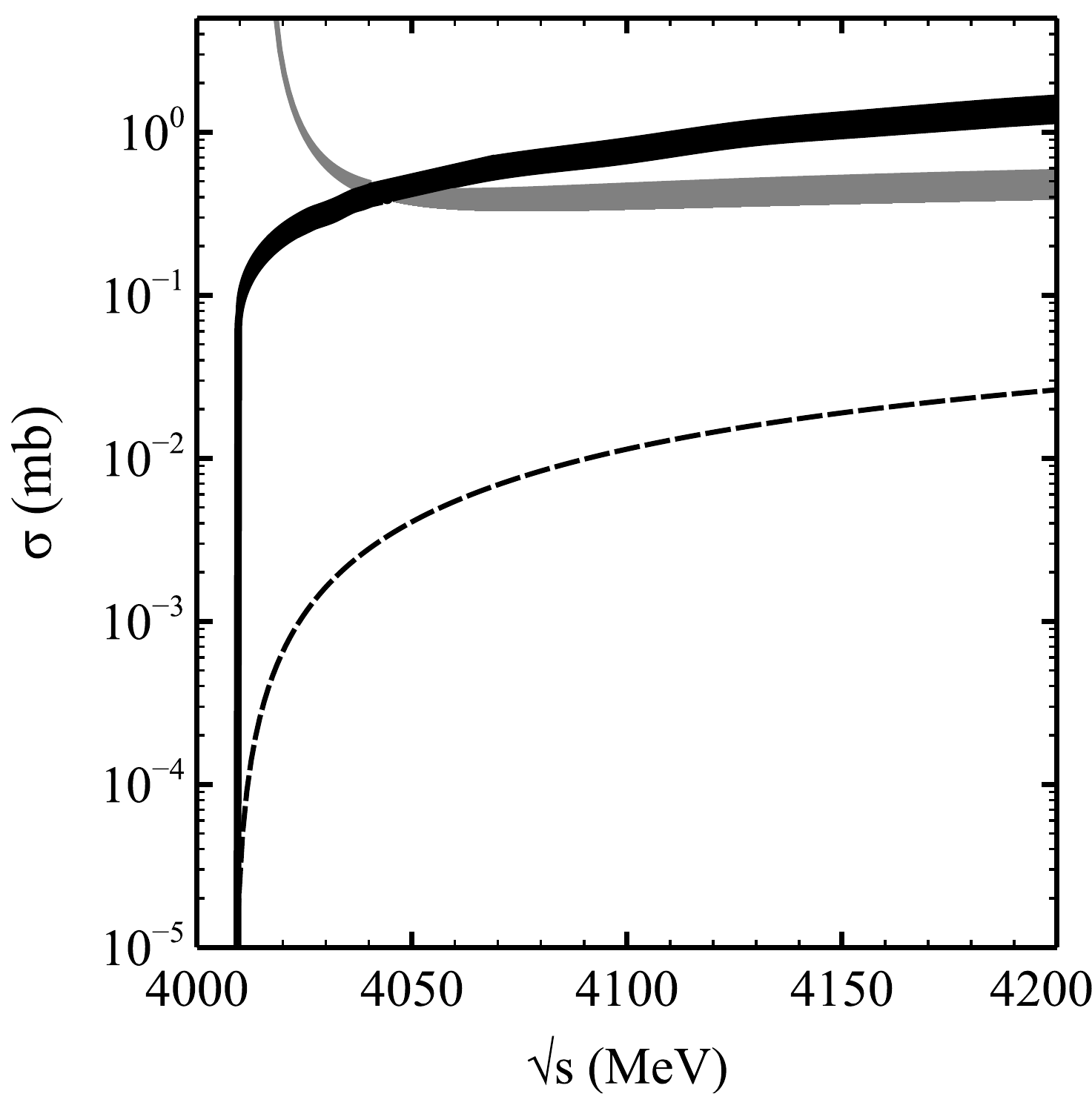}
\caption{Cross sections for the different reactions studied. The dashed line, dark shaded region and light shaded region represent the total cross section for the $\bar D D\to \pi X$, $\bar D^* D\to\pi X$ and $\bar D^* D^*\to\pi X$ reactions, respectively.}\label{crossall}
\end{figure}

In Fig.~\ref{crossall} we show a comparison of the total cross sections obtained for the three reactions studied in this paper. The dashed-line is the result for the $\bar D D\to \pi X$ reaction, while the shaded
areas correspond to the cross sections for the processes $\bar D^* D\to\pi X$ (dark shaded region) and $\bar D^* D^*\to\pi X$ (light shaded region). As can be seen,  the cross section for the
$\bar D^* D\to \pi X$ process exceeds the one of $\bar D^* D^*\to\pi X$ when increasing the energy.

\subsection{Inclusion of Form Factors}
Finally, it should be mentioned that we could have also included form factors in the vertices when evaluating the cross sections for the processes studied in this paper. In Ref.~\cite{Cho3} monopole form factors
of the type
\begin{align}
\frac{\Lambda^2}{\Lambda^2+\vec{q}^{\,\,2}},\label{FF}
\end{align}
were considered in the calculation of the cross sections for each of the vertices involving a $t$ or $u$ channel exchange of a heavy meson, with $\Lambda=2000$ MeV and $\vec{q}$ the momentum transfer in the CM frame. This would result in a change of the magnitude for these cross sections, specially at higher energies.

\begin{figure}
\includegraphics[width=0.5\textwidth]{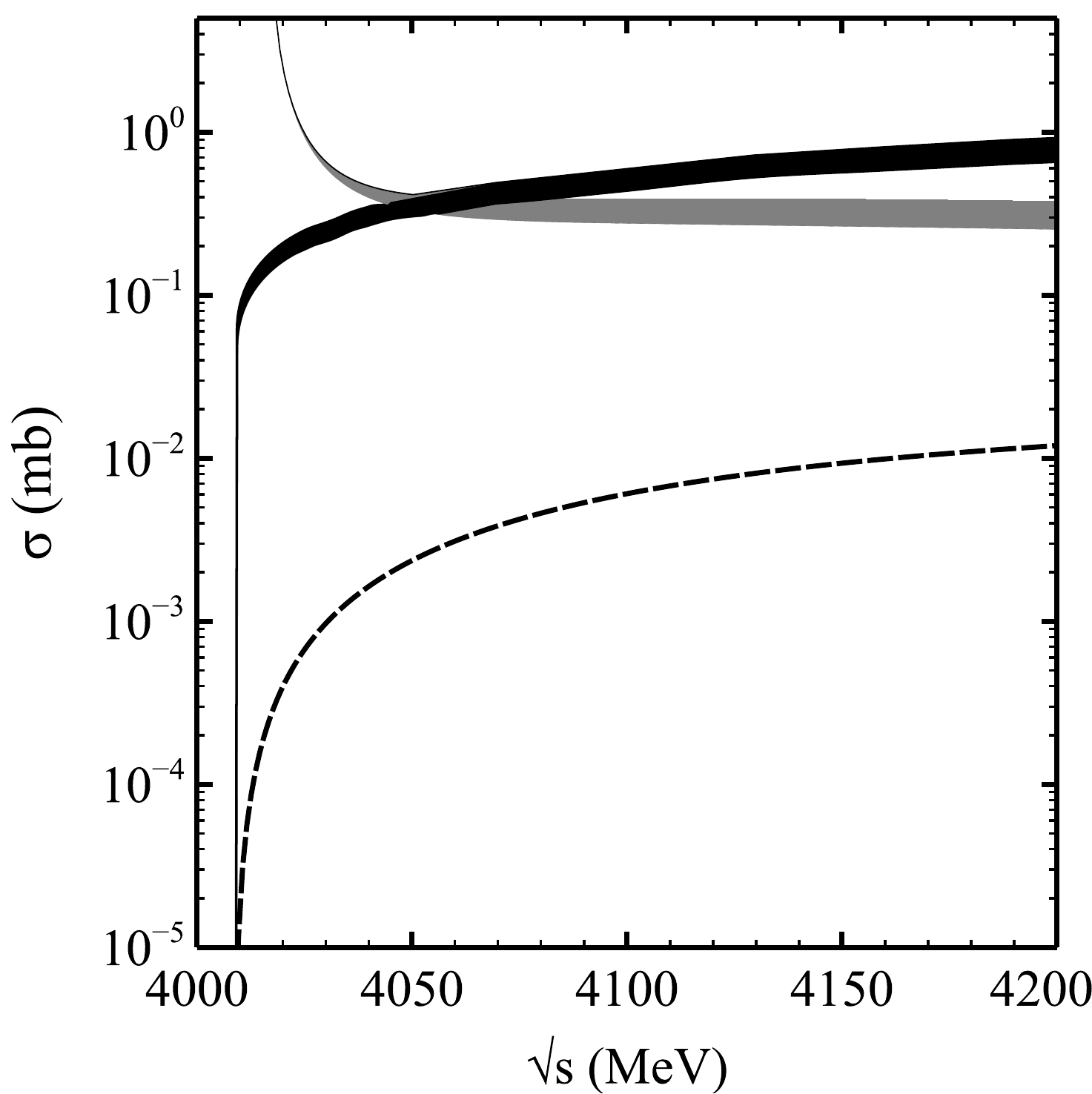}
\caption{Cross sections for the different reactions studied using form factors. The dashed line, dark shaded region and light 
shaded region correspond to the total cross section for the $\bar D D\to \pi X$, $\bar D^* D\to\pi X$ and $\bar D^* D^*\to\pi X$ 
reactions, respectively.}
\label{allFF}
\end{figure}

In Fig.~\ref{allFF} we show the cross sections for the different reactions studied here when we take into account the inclusion of the form factors of Eq.~(\ref{FF}). As can be seen, a reduction in the cross sections of around a factor 2 is found at an energy of 200 MeV above the threshold. This reduction is similar to the one found  in Ref.~\cite{Oh} for the $\pi J\psi$ absorption cross sections with the form
factor of Eq.~(\ref{FF}).

\section{Summary}\label{Sum}

In this work we have obtained the production cross sections of the reactions $\bar D D \to\pi X$, $\bar D^* D \to\pi X$ and   $\bar D^* D^*\to\pi X$, considering $X(3872)$ as a molecular state of $\bar D D^*-\textrm{c.c}$. 
We have shown that the consideration of the neutral as well as the charged hadrons coupling to $X$ is important for the 
evaluation of the cross sections. Next, to obtain the cross section for the process $\bar D^* D\to\pi X$ we have included the 
contribution of the anomalous vertex $X\bar D^* D^*$. With this result, we have estimated the $X\bar D^* D^*$ coupling and used 
it to calculate the cross section for the reaction $\bar D^* D^*\to\pi X$. The contribution to the cross section from the 
vertex $X\bar D^* D^*$ turns out to be important and could play an important role in the determination of the abundance of 
the $X$ meson in heavy ion collisions.

Our results, specially those presented in Fig. \ref{allFF}, pave the way for a new round of calculations of $X$ abundancies in a 
hadron gas, as outlined in Ref. \cite{Cho3}. With them we can compute the average cross sections $<\sigma_{a b \to c d} v_{ab}>$, 
where $v_{ab}$ is the relative velocity between the colliding particles and the brackets denote the average over the thermal 
distributions of the incoming particles $a$ and $b$. Knowing $<\sigma_{a b \to c d} v_{ab}>$ and the inverse cross sections 
(obtained through detailed balance relations), we can solve the kinetic equations and obtain the abundancies as a function of time. 
This requires some modeling of the quark gluon plasma and we postpone these calculations for a future work. We emphasize that 
we expect to find some significant differences with respect to the results found in Ref.~\cite{Cho3}, because the processes 
 $\bar D D \to\pi X$ and  $\bar D^* D^* \to\pi X$ have been recalculated and, more importantly, the process 
$\bar D^* D \to\pi X$ has been included. This latter was found to give the most important contribution of all the three  
processes considered.

In Ref.~\cite{Cho3} the authors suggested that the measurement of the $X$ multiplicity would be very useful to determine 
its structure. Molecular $\bar D D^*$ states were predicted to have a multiplicity 18 times bigger than the 
tetraquarks states. Therefore, just by measuring the number of produced $X$'s we would be able to know whether it is a 
meson molecule or a  tetraquark. It will be very interesting to see what will happen to this prediction after the 
correction in the cross sections.

Finally it is important to mention that the predictions discussed here will eventually be tested in the laboratory. 
In the  near future, with the implementation of the heavy flavor tracker in the STAR experiment, 
we will be  able to find charmed mesons  coming from the X(3872) mesons
and measure the yield of X(3872) mesons produced by the coalescence in heavy ion collisions.

\section{Acknowledgements}
We thank professor Eulogio Oset for very useful discussions. The authors would like to thank the Brazilian funding agencies FAPESP and CNPq for the financial support. 

\appendix
\section{Coefficients needed in the evaluation of the $u$-channel amplitudes for the diagrams in Fig.~\ref{blob}}\label{AppenA}
In this appendix we list the different isospin coefficients needed to determine the $u$-channel amplitudes associated with the diagrams in Fig.~\ref{blob}. These coefficients are actually the product of the different isospin coefficients at the vertices in the diagrams shown in Fig.~\ref{blob}. Tables~\ref{ApenF3a} and \ref{ApenF3b} list the internal hadrons considered. The amplitudes associated with the diagrams in Fig.~\ref{blob} are calculated for each of these internal hadrons and summed up eventually, as explained in Sec.~\ref{For}. 

\begin{table}
\caption{Coefficients $F^{(Q_{1i},Q_{2i})}_A=-F^{(Q_{1i},Q_{2i})}_C$ appearing in Eqs.~(\ref{Ua}) and~(\ref{Uc}) and which are associated with the amplitudes of the diagrams in Figs.~\ref{blob}a~and~\ref{blob}c. The pair $(Q_{1i},Q_{2i})$ denotes the charge of the particles forming the initial state of the reaction [as convention, $Q_{1i}$ ($Q_{2i}$) is the charge of the particle with charm $-1$ ($+1$)].}\label{ApenF3a}
\begin{tabular}{cccccc}
$(Q_{1i},Q_{2i})$&$P$&$\bar P_X$&$V_X$&$D^*$\quad&$F_A$\\
\hline
\multirow{6}{*}{$(0,0)$}\quad\quad&$\pi^0$&$\bar D^0$&$D^{*0}$&$D^{*0}$\quad\quad&$-1/2\sqrt{2}$\\
\quad\quad&$\eta$&$\bar D^0$&$D^{*0}$&$D^{*0}$\quad\quad&$-1/3\sqrt{2}$\\
\quad\quad&$\eta^\prime$&$\bar D^0$&$D^{*0}$&$D^{*0}$\quad\quad&$-1/6\sqrt{2}$\\
\quad\quad&$\eta_c$&$\bar D^0$&$D^{*0}$&$D^{*0}$\quad\quad&$1/\sqrt{2}$\\
\quad\quad&$\pi^+$&$D^-$&$D^{*+}$&$D^{*0}$\quad\quad&$-1/\sqrt{2}$\\
\quad\quad&$K^+$&$D^-_s$&$D^{*+}_s$&$D^{*0}$\quad\quad&$-1/\sqrt{2}$\\
\hline
\multirow{6}{*}{$(-,+)$}\quad\quad&$\pi^0$&$D^-$&$D^{*+}$&$D^{*+}$\quad\quad&$1/2\sqrt{2}$\\
\quad\quad&$\eta$&$D^-$&$D^{*+}$&$D^{*+}$\quad\quad&$1/3\sqrt{2}$\\
\quad\quad&$\eta^\prime$&$D^-$&$D^{*+}$&$D^{*+}$\quad\quad&$1/6\sqrt{2}$\\
\quad\quad&$\eta_c$&$D^-$&$D^{*+}$&$D^{*+}$\quad\quad&$-1/\sqrt{2}$\\
\quad\quad&$\pi^-$&$\bar D^0$&$D^{*0}$&$D^{*+}$\quad\quad&$1/\sqrt{2}$\\
\quad\quad&$K^0$&$D^-_s$&$D^{*+}_s$&$D^{*+}$\quad\quad&$1/\sqrt{2}$\\
\hline
\multirow{6}{*}{$(-,0)$}\quad\quad&$\pi^0$&$D^-$&$D^{*+}$&$D^{*+}$\quad\quad&$-1/2$\\
\quad\quad&$\eta$&$D^-$&$D^{*+}$&$D^{*+}$\quad\quad&$-1/3$\\
\quad\quad&$\eta^\prime$&$D^-$&$D^{*+}$&$D^{*+}$\quad\quad&$-1/6$\\
\quad\quad&$\eta_c$&$D^-$&$D^{*+}$&$D^{*+}$\quad\quad&$1$\\
\quad\quad&$\pi^-$&$\bar D^0$&$D^{*0}$&$D^{*+}$\quad\quad&$-1$\\
\quad\quad&$K^0$&$D^-_s$&$D^{*+}_s$&$D^{*+}$\quad\quad&$-1$\\
\hline
\multirow{6}{*}{$(0,+)$}\quad\quad&$\pi^0$&$\bar D^0$&$D^{*0}$&$D^{*0}$\quad\quad&$-1/2$\\
\quad\quad&$\eta$&$\bar D^0$&$D^{*0}$&$D^{*0}$\quad\quad&$-1/3$\\
\quad\quad&$\eta^\prime$&$\bar D^0$&$D^{*0}$&$D^{*0}$\quad\quad&$-1/6$\\
\quad\quad&$\eta_c$&$\bar D^0$&$D^{*0}$&$D^{*0}$\quad\quad&$1$\\
\quad\quad&$\pi^+$&$D^-$&$D^{*+}$&$D^{*0}$\quad\quad&$-1$\\
\quad\quad&$K^+$&$D^-_s$&$D^{*+}_s$&$D^{*0}$\quad\quad&$-1$\\
\end{tabular}
\end{table}

\begin{table}
\caption{Coefficients $F^{(Q_{1i},Q_{2i})}_B=F^{(Q_{1i},Q_{2i})}_D$ appearing in Eqs.~(\ref{Ub})~and~(\ref{Ud}) and which are associated with the amplitudes of the diagrams in Figs.~\ref{blob}b~and~\ref{blob}d. The pair $(Q_{1i},Q_{2i})$ denotes the charge of the particles forming the initial state of the reaction [as convention, $Q_{1i}$ ($Q_{2i}$) is the charge of the particle with charm $-1$ ($+1$)].}\label{ApenF3b}
\begin{tabular}{cccccc}
$(Q_{1i},Q_{2i})$&$P$&$\bar P_X$&$V_X$&$D^*$\quad&$F_B$\\
\hline
\multirow{6}{*}{$(0,0)$}\quad\quad&$\rho^0$&$\bar D^0$&$D^{*0}$&$D^{*0}$\quad\quad&$1/2\sqrt{2}$\\
\quad\quad&$\omega$&$\bar D^0$&$D^{*0}$&$D^{*0}$\quad\quad&$1/2\sqrt{2}$\\
\quad\quad&$\phi$&$\bar D^0$&$D^{*0}$&$D^{*0}$\quad\quad&$0$\\
\quad\quad&$J/\psi$&$\bar D^0$&$D^{*0}$&$D^{*0}$\quad\quad&$-1/\sqrt{2}$\\
\quad\quad&$\rho^+$&$D^-$&$D^{*+}$&$D^{*0}$\quad\quad&$1/\sqrt{2}$\\
\quad\quad&$K^{*+}$&$D^-_s$&$D^{*+}_s$&$D^{*0}$\quad\quad&$1/\sqrt{2}$\\
\hline
\multirow{6}{*}{$(-,+)$}\quad\quad&$\rho^0$&$D^-$&$D^{*+}$&$D^{*+}$\quad\quad&$-1/2\sqrt{2}$\\
\quad\quad&$\omega$&$D^-$&$D^{*+}$&$D^{*+}$\quad\quad&$-1/2\sqrt{2}$\\
\quad\quad&$\phi$&$D^-$&$D^{*+}$&$D^{*+}$\quad\quad&$0$\\
\quad\quad&$J/\psi$&$D^-$&$D^{*+}$&$D^{*+}$\quad\quad&$1/\sqrt{2}$\\
\quad\quad&$\rho^-$&$\bar D^0$&$D^{*0}$&$D^{*+}$\quad\quad&$-1/\sqrt{2}$\\
\quad\quad&$K^{*0}$&$D^-_s$&$D^{*+}_s$&$D^{*+}$\quad\quad&$-1/\sqrt{2}$\\
\hline
\multirow{6}{*}{$(-,0)$}\quad\quad&$\rho^0$&$D^-$&$D^{*+}$&$D^{*+}$\quad\quad&$1/2$\\
\quad\quad&$\omega$&$D^-$&$D^{*+}$&$D^{*+}$\quad\quad&$1/2$\\
\quad\quad&$\phi$&$D^-$&$D^{*+}$&$D^{*+}$\quad\quad&$0$\\
\quad\quad&$J/\psi$&$D^-$&$D^{*+}$&$D^{*+}$\quad\quad&$-1$\\
\quad\quad&$\rho^-$&$\bar D^0$&$D^{*0}$&$D^{*+}$\quad\quad&$1$\\
\quad\quad&$K^{*0}$&$D^-_s$&$D^{*+}_s$&$D^{*+}$\quad\quad&$1$\\
\hline
\multirow{6}{*}{$(0,+)$}\quad\quad&$\rho^0$&$\bar D^0$&$D^{*0}$&$D^{*0}$\quad\quad&$1/2$\\
\quad\quad&$\omega$&$\bar D^0$&$D^{*0}$&$D^{*0}$\quad\quad&$1/2$\\
\quad\quad&$\phi$&$\bar D^0$&$D^{*0}$&$D^{*0}$\quad\quad&$0$\\
\quad\quad&$J/\psi$&$\bar D^0$&$D^{*0}$&$D^{*0}$\quad\quad&$-1$\\
\quad\quad&$\rho^+$&$D^-$&$D^{*+}$&$D^{*0}$\quad\quad&$1$\\
\quad\quad&$K^{*+}$&$D^-_s$&$D^{*+}_s$&$D^{*0}$\quad\quad&$1$\\
\end{tabular}
\end{table}

\section{Evaluation of the integrals related to the diagrams in Fig.~\ref{blob}}\label{AppenB}
Using Lorentz covariance, the integral in Eq.~(\ref{int}) can be written as
\begin{align}
\mathcal{I}_{\mu\alpha^\prime}=i(a_A\,g_{\mu \alpha^\prime}+b_A\,p_{1\mu}\,p_{1\alpha^\prime}+c_A\,p_{1\alpha^\prime}\,p_{4\mu}+d_A\,p_{1\mu}\,p_{4\alpha^\prime}+e_A\,p_{4\mu}\,p_{4\alpha^\prime}),\label{cov}
\end{align}
and considering the Lorentz gauge $p\cdot\epsilon(p)=0$ we can write the expression
\begin{align}
\epsilon^\mu_{\bar D^*} (p_1)\epsilon^{\mu^\prime\nu^\prime\alpha^\prime\beta^\prime}\mathcal{I}_{\mu\alpha^\prime}\epsilon_{X\beta^\prime}(p_4),
\end{align}
present in Eq.~(\ref{Ua}) as
\begin{align}
\epsilon^\mu_{\bar D^*} (p_1)\epsilon^{\mu^\prime\nu^\prime\alpha^\prime\beta^\prime}\mathcal{I}_{\mu\alpha^\prime}\epsilon_{X\beta^\prime}(p_4)=\epsilon^\mu_{\bar D^*} (p_1)\epsilon^{\mu^\prime\nu^\prime\alpha^\prime\beta^\prime}i(a_A\,g_{\mu \alpha^\prime}+c_A\,p_{1\alpha^\prime}+e_A\,p_{4\mu}\,p_{4\alpha^\prime})\epsilon_{X\beta^\prime}(p_4),\label{simp}
\end{align}
and, thus, only the coefficients $a_A$, $c_A$ and $e_A$ of Eq.~(\ref{cov}) need to be calculated. To do this, we make use of the Feynman parametrization and write
\begin{align}
\frac{1}{\alpha \beta\gamma}=2\int_0^1 dx\int_0^x dy \frac{1}{\left[\alpha+(\beta-\alpha)x+(\gamma-\beta)y\right]^3},\label{fey}
\end{align}
where 
\begin{align}
\alpha&\equiv(p_1-k)^2-m^2_P,\nonumber\\
\beta&\equiv k^2-m^2_{\bar P_X},\\
\gamma&\equiv (p_4-k)^2-m^2_{V_X}.\nonumber
\end{align}
In this way, 
\begin{align}
\left[\alpha+(\beta-\alpha)x+(\gamma-\beta)y\right]={k^\prime}^2+r_1,\label{conv}
\end{align}
where we have defined
\begin{align}
k^\prime\equiv k+p_1(x-1)-p_4 y,\label{kp}
\end{align}
\begin{align}
r_1\equiv (x-1)(-m^2_{\bar D^*}x+m^2_P+2\,p_1\cdot p_4\, y)-y[m^2_X(y-1)+m^2_{V_X}]-m^2_{\bar P_X}(x-y).
\end{align}
Using Eqs.~(\ref{fey}),~(\ref{conv}) and~(\ref{kp}) in Eq.~(\ref{int}), we can identify the coefficients $a_A$, $c_A$ and $e_A$ of Eq.~(\ref{simp}) as: 
\begin{align}
i\,a_A&=-\int_0^1 dx\int_0^x dy\int\frac{d^4k^\prime}{(2\pi)^4}\frac{{k^\prime}^2}{({k^\prime}^2+r_1+i\epsilon)^3},\label{acoef}\\
i\, c_A&=4\int_0^1dx\,(x-1)\int_0^xdy\,y\int\frac{d^4k^\prime}{(2\pi)^4}\frac{1}{({k^\prime}^2+r_1+i\epsilon)^3},\label{ccoef}\\
i\, e_A&=-4\int_0^1dx\int_0^xdy\,y(y-1)\int\frac{d^4k^\prime}{(2\pi)^4}\frac{1}{({k^\prime}^2+r_1+i\epsilon)^3}\label{ecoef}.
\end{align}
Using the relation
\begin{align}
\int\frac{d^4k^\prime}{(2\pi)^4}\frac{1}{({k^\prime}^2+r_1+i\epsilon)^3}=\frac{i}{2^5\pi^2(r_1+i\epsilon)},\label{intconverg}
\end{align}
we can reduce Eqs.~(\ref{ccoef})~and~(\ref{ecoef}) to,
\begin{align}
c_A&=\frac{1}{2^3\pi^2}\int_0^1dx\,(x-1)\int_0^xdy\,\frac{y}{r_1+i\epsilon},\\
e_A&=-\frac{1}{2^3\pi^2}\int_0^1dx\int_0^xdy\,\frac{y(y-1)}{r_1+i\epsilon}.
\end{align}
The determination of the coefficient $a_A$ is more complicated since the integral in the variable $k^\prime$ present in Eq.~(\ref{acoef}) is logaritmicaly divergent. In fact, the calculation of this coefficient is simpler if we
write the integral in terms of the $k$ variable and not in terms of $k^\prime$. Using Eqs.~(\ref{fey}),~(\ref{kp}) and~(\ref{intconverg}) we can write
\begin{align}
a_A=\frac{1}{2}\mathcal{F}(m_{\bar P_X},m_P,m_{V_X})+\frac{1}{2^5\pi^2}\int_0^1 dx\int_0^x dy\frac{m^2_{\bar D^*}(x-1)^2+m^2_Xy^2-2\,p_1\cdot p_4\,(x-1)y}{r_1+i\epsilon},
\end{align}
where we have defined
\begin{align}
\mathcal{F}(m_1,m_2,m_3)\equiv i\int\frac{d^4k}{(2\pi)^4}\frac{k^2}{[k^2-m^2_1+i\epsilon][(p_1-k)^2-m^2_2+i\epsilon][(p_4-k)^2-m^2_3+i\epsilon]}.\label{Fint}
\end{align}
To determine the integral in Eq.~(\ref{Fint}) we use first Cauchy's theorem to perform the integration of the temporal part of the $k$ variable, finding
\begin{align}
\mathcal{F}(m_1,m_2,m_3)=\int\frac{d^3 k}{(2\pi)^3}\frac{\mathcal{N}(m_1,m_2,m_3)}{\mathcal{D}(m_1,m_2,m_3)},\label{Fm1m2m3}
\end{align}
with,
\begin{align}
\mathcal{N}(m_1,m_2,m_3)&=|\vec{k}|^2\Big[(p^0_1)^2\omega_2(\omega_1+\omega_3)-2\,p^0_1\,p^0_4\,\omega_2\,\omega_3\nonumber\\
&\quad-(\omega_1+\omega_2)\left\{(\omega_1+\omega_3)(\omega_2+\omega_3)(\omega_1+\omega_2+\omega_3)-(p^0_4)^2\omega_3\right\}\Big]\nonumber\\
&\quad+\omega_1\Big[(p^0_1)^2\left\{\omega_3(\omega_1+\omega_3)(\omega_1+\omega_2+\omega_3)-(p^0_4)^2(\omega_2+\omega_3)\right\}\nonumber\\
&\quad+2\,p^0_1\,p^0_4\,\omega_1\,\omega_2\,\omega_3-\omega_2(\omega_1+\omega_2)\left\{\omega_3(\omega_1+\omega_3)(\omega_2+\omega_3)\right.\nonumber\\
&\quad\left.-(p^0_4)^2(\omega_1+\omega_2+\omega_3)\right\}\Big],
\end{align}
\begin{align}
\mathcal{D}(m_1,m_2,m_3)&=-2\,\omega_1\,\omega_2\,\omega_3\Big[(p^0_1)^2-(\omega_1+\omega_2)^2+i\epsilon\Big]\Big[(p^0_4)^2-(\omega_1+\omega_3)^2+i\epsilon\Big]\nonumber\\
&\quad\times\Big[(p^0_4-p^0_1)^2-(\omega_2+\omega_3)^2+i\epsilon\Big].
\end{align}
and where we have defined
\begin{align}
\omega_1&=\sqrt{|\vec{k}|^2+m^2_1},\nonumber\\
\omega_2&=\sqrt{(\vec{k}-\vec{p}_1)^2+m^2_2},\\
\omega_3&=\sqrt{(\vec{k}-\vec{p}_4)^2+m^2_3}.\nonumber
\end{align}
The quantities $p^0_1$ and $p^0_4$ correspond to the center of mass energies of the externals $\bar D^*$ and $X$. The integration on the variable $|\vec{k}|$ in Eq.~(\ref{Fm1m2m3}) is logarithmically divergent and it can be regularized with a cut-off of a natural size, of the order of 1 GeV~\cite{Nagahiro1,Nagahiro2,Raquel}.

Here a comment is in order. In the determination of the residues of Eq.~(\ref{Fint}) we encounter terms with undefined polarizations, of the type
\begin{align}
\frac{1}{W-W_0-i\epsilon+i\epsilon^\prime}\,,
\end{align}
with $W$ and $W_0$ being linear combinations of energy type variables ($\omega_1$, $\omega_2$, $p^0_1$, etc.). These kind of terms have been often referred in the literature as ``fallacious poles"~\cite{Raquel,Javi}. It is interesting to notice that the sum
of the different residues, which contains these type of terms, is such that the terms with undefined polarization can be factorized in the resulting numerator and, thus, get cancelled with the ones present in the denominator, removing in this way any kind of ambiguity.

The determination of the rest of the integrals needed to determine the amplitudes in the diagrams of Fig.~\ref{blob} is analogous to the one we just saw. In the following we just list some definitions and the results
for these integrals. 

For the integrals in Eqs.~(\ref{HJR}) we have
\begin{align}
\mathcal{H}^\sigma_\alpha&=i(a_{1B}\,g^\sigma_\alpha+d_{1B}\,p_{4\alpha}\,p_1^\sigma),\\
\mathcal{J}_{\alpha\nu^\prime}&=i(a_{2B}\,g_{\alpha\nu^\prime}+d_{2B}\,p_{4\alpha}\,p_{1\nu^\prime}+e_{2B}\,p_{4\alpha}\,p_{4\nu^\prime}),
\end{align}
where we have omitted terms which are zero after contracting these integrals with the Levi-Civita tensor present in Eq.~(\ref{Ub}) or after using the Lorentz condition $p\cdot\epsilon(p)=0$ and
\begin{align}
a_{1B}&=\frac{1}{4}\mathcal{F}(m_{\bar P_X},m_V,m_{V_X})+\frac{1}{2^6\pi^2}\int_0^1dx\int_0^xdy\frac{m^2_{\bar D^*}(x-1)^2+m^2_Xy^2-2\,p_1\cdot p_4(x-1)y}{r_2+i\epsilon},\nonumber
\end{align}
\begin{align}
d_{1B}&=\frac{1}{2^4\pi^2}\int_0^1dx\,(x+1)\int_0^x dy \frac{y}{r_2+i\epsilon},\nonumber\\
a_{2B}&=-2a_{1B},\nonumber\\
d_{2B}&=-\frac{1}{2^4\pi^2}\int_0^1dx\,(2x-1)\int_0^xdy\frac{y}{r_2+i\epsilon},\nonumber\\
e_{2B}&=\frac{1}{2^4\pi^2}\int_0^1 dx\int_0^x dy \frac{y(2y-1)}{r_2+i\epsilon},\nonumber
\end{align}
\begin{align}
r_2=(x-1)(-m^2_{\bar D^*}x+m^2_V+2\, p_1\cdot p_4\, y)-y\,[m^2_X(y-1)+m^2_{V_X}]-m^2_{\bar P_X}(x-y).\nonumber
\end{align}
In case of the integral in Eq.~(\ref{R}), we have
\begin{align}
\mathcal{R}_{\alpha\mu^\prime}=i(a_C\,g_{\alpha\mu^\prime}+d_C\,p_{4\alpha}\,p_{1\mu^\prime}+e_C\,p_{4\alpha}\,p_{4\mu^\prime})\label{R2}
\end{align}
with,
\begin{align}
a_{C}&=\frac{1}{2}\mathcal{F}(m_{\bar V_X},m_P,m_{P_X})+\frac{1}{2^5\pi^2}\int_0^1dx\int_0^xdy\frac{m^2_{\bar D^*}(x-1)^2+m^2_Xy^2-2\,p_1\cdot p_4(x-1)y}{r_3+i\epsilon},\nonumber
\end{align}
\begin{align}
d_C&=\frac{1}{2^3\pi^2}\int_0^1dx\,(x-1)\int_0^xdy \frac{y}{r_3+i\epsilon},\nonumber\\
e_C&=-\frac{1}{2^3\pi^2}\int_0^1dx\int_0^xdy\,\frac{y^2}{r_3+i\epsilon},\nonumber
\end{align}
\begin{align}
r_3=(x-1)(-m^2_{\bar D^*}x+m^2_P+2\,p_1\cdot p_4\,y)-y[m^2_X(y-1)+m^2_{P_X}]-m^2_{\bar V_X}(x-y),\nonumber
\end{align}
and, as done before, we omit terms in Eq.~(\ref{R2}) which give zero contribution due to the antisymmetric properties of the Levi-Civita tensor present in Eq.~(\ref{Uc}) or due to the Lorentz condition.

Similarly, for the integrals in Eqs.~(\ref{Q})~and~(\ref{S})
\begin{align}
\mathcal{Q}_{\alpha^\prime\beta^\prime}&=i(a_{1D}\,g_{\alpha^\prime\beta^\prime}+c_{1D}\,p_{1\alpha^\prime}\,p_{4\beta^\prime}+d_{1D}\,p_{4\alpha^\prime}\,p_{1\beta^\prime}),\nonumber\\
\mathcal{Q}_{\alpha^\prime\nu}&=i(a_{2D}\,g_{\alpha^\prime\nu}+c_{2D}\,p_{1\alpha^\prime}\,p_{4\nu}+e_{2D}\,p_{4\alpha^\prime}\,p_{4\nu}),\label{QA}\\
\mathcal{S}_{\alpha^\prime\sigma}&=i(a_{3D}\,g_{\alpha^\prime\sigma}+b_{3D}\,p_{1\alpha^\prime}\,p_{1\sigma}+d_{3D}\,p_{4\alpha^\prime}\,p_{1\sigma}),\nonumber
\end{align}
with,
\begin{align}
a_{1D}&=\frac{1}{4}\mathcal{F}(m_{\bar V_X},m_V,m_{P_X})+\frac{1}{2^6\pi^2}\int_0^1dx\int_0^xdy\frac{m^2_{\bar D^*}(x-1)^2+m^2_Xy^2-2\,p_1\cdot p_4(x-1)y}{r_4+i\epsilon},\nonumber
\end{align}
\begin{align}
c_{1D}&=\frac{1}{2^4\pi^2}\int_0^1dx\,x\int_0^xdy\,\frac{y}{r_4+i\epsilon},\nonumber\\
d_{1D}&=\frac{1}{2^4\pi^2}\int_0^1dx\,(x-1)\int_0^xdy\frac{y}{r_4+i\epsilon},\nonumber\\
a_{2D}&=a_{1D},\nonumber\\
c_{2D}&=c_{1D},\nonumber\\
e_{2D}&=-\frac{1}{2^4\pi^2}\int_0^1dx\int_0^xdy\frac{y^2}{r_4+i\epsilon},\nonumber\\
a_{3D}&=-a_{1D},\nonumber\\
b_{3D}&=\frac{1}{2^4\pi^2}\int_0^1dx\,x(x+1)\int_0^xdy\frac{1}{r_4+i\epsilon},\nonumber\\
d_{3D}&=-\frac{1}{2^4\pi^2}\int_0^1dx (x+1)\int_0^xdy\frac{y}{r_4+i\epsilon},\nonumber
\end{align}
\begin{align}
r_4=(x-1)(-m^2_{\bar D^*}x+m^2_V+2\,p_1\cdot p_4\,y)-y[m^2_X(y-1)+m^2_{P_X}]-m^2_{\bar V_X}(x-y).\nonumber
\end{align}
In Eqs.~(\ref{QA}) we have omitted terms which are zero due to the antisymmetric properties of the Levi-Civita tensor present in Eq.~(\ref{Ud}) or the Lorentz condition.

\end{document}